\documentclass[sn-mathphys]{sn-jnl}% Math and Physical Sciences Reference Style
\usepackage[utf8]{inputenc}
\jyear{2022}%

\theoremstyle{thmstyleone}%
\theoremstyle{thmstyletwo}%

\theoremstyle{thmstylethree}%

\newcommand{\lsim}{\,\raise 0.4ex\hbox{$<$}\kern -0.8em\lower 0.62ex\hbox{$\sim$}\,}
\newcommand{\gsim}{\,\raise 0.4ex\hbox{$>$}\kern -0.7em\lower 0.62ex\hbox{$\sim$}\,}
\newcommand{\be}{\begin{equation}}
\newcommand{\ee}{\end{equation}}
\newcommand{\bea}{\begin{eqnarray}}
\newcommand{\eea}{\end{eqnarray}}
\newcommand{\bean}{\begin{eqnarray*}}
\newcommand{\eean}{\end{eqnarray*}}
\newcommand{\dd}{\partial}
\newcommand{\bx}{ {\bf x}}

\newcommand{\bV}{{\bf V}}

\newcommand{\bn}{{\bf n}}

\newcommand{\bfe}{{\bf e}}

\newcommand{\HH}{\mathcal{H}}

\newcommand{\al}{\alpha}

\newcommand{\de}{\delta}
\newcommand{\De}{\Delta}

\newcommand{\ga}{\gamma}

\newcommand{\ka}{\kappa}

\newcommand{\La}{\Lambda}
\newcommand{\la}{\lambda}
\newcommand{\Om}{\Omega}

\newcommand{\si}{\sigma}

\newcommand{\bnabla}{\ensuremath{\mathbf{\nabla}}}%connection sur variete

\begin{document}

\title[Testing General Relativity with LSS]{Testing General Relativity with Cosmological Large Scale Structure}


\author*{\fnm{Ruth} \sur{Durrer}}\email{ruth.durrer@unige.ch}


\affil{\orgdiv{Department of Theoretical Physics}, \orgname{Université de Genève}, \orgaddress{\street{Quai E. Ansermet 24 }, \city{Genève}, \postcode{1211}, \country{Switzerland}}}

%\affil[2]{\orgdiv{Department}, \orgname{Organization}, \orgaddress{\street{Street}, \city{City}, \postcode{10587}, \state{State}, \country{Country}}}

\

\abstract{In this paper  I investigate the possibility to test Einstein's equations with  observations of cosmological large scale structure. I first show that we have not tested the equations in observations concerning only the homogeneous and isotropic Universe. I then show with several examples how we can do better when considering the fluctuations of both, the energy momentum tensor and the metric. This is illustrated with galaxy number counts, intensity mapping and cosmic shear, three  examples that are by no means exhaustive.  }




\maketitle

\section{Introduction}\label{sec1}
This contribution is written to honour my dear colleague  Thanu Padmanabhan (Paddy) who passed away on September 17, 2021. I have great admiration for Paddy, especially for his crystal clear arguments which are found in all his papers and which are especially present in his books.  His three volumes on Astrophysics~\cite{astro1,astro2,astro3} and his book on General Relativity~\cite{rela} are master pieces of textbooks in our field. 

Paddy's excellence in teaching is a consequence of his deep understanding of physics.
This is also reflected in his choice of research topics which are in most cases an attempt to address a really deep open question in theoretical physics, like e.g. the riddle of dark energy.

In this article written for him I also would like to discuss an important 	question, namely : \\
\begin{center}Can we test the non-vacuum sector of General Relativity ? \vspace{10pt}
\end{center}
The vacuum equations,
\be
R_{\mu\nu} =0\,,
\ee
have been tested in many ways: First via light deflection which made Einstein famous, but also via  the perihelion advance of Mercury. Both these observations test the Schwarzschild exterior  solution of stars. Later, especially binary pulsars allowed very detailed tests of the most important vacuum solutions, the Schwarzschild and Kerr spacetimes. We have also measured the Lense-Thirring effect and, finally  gravitational waves from binary black holes or neutron stars have been discovered. Even though the interior of a neutron star is not a vacuum solution, this has so far not been tested with significant precision. The same is true for the interior of  other stellar objects where relativistic effects are less relevant  than in neutron stars.

But can we also test the equations in the presence of matter,
\be 
G_{\mu\nu} = 8\pi GT_{\mu\nu}\,.
\ee
Here we include a possible cosmological constant $\La$ as vacuum energy given by $\rho_v = \La/(8\pi G)$ in the energy momentum tensor. Since there is no experiment that can distinguish between vacuum energy and a cosmological constant, we should not do so  also in our theories. Therefore we use the terms 'cosmological constant' and 'vacuum energy' as synonyms. 
To test Einstein's equations we must  measure both, the geometry of spacetime that determines the Einstein tensor $G_{\mu\nu}$ and the energy momentum distribution in the Universe that determines $T_{\mu\nu}$. In this article I shall argue that we can do this in the near future in cosmology. The cosmological metric is probably the most relevant and best measured non-vacuum solution of General Relativity  (GR).

The reminder of this article is structured as follows. In the next section I discuss homogeneous cosmology and argue that we have not tested Einstein's equations in observations of the background metric. In Section~\ref{sec:LSS} we  discuss the observations of cosmological large scale structure and I  propose several ways to test Einstein's equations with these observations.
Is Section~\ref{sec:con} I summarize and conclude.

\section{Homogeneous cosmology}\label{sec:hom}
The most relevant known non-vacuum solution which is being measured with more and more accurate observations is the cosmological solution. At very large scales, we suppose that the geometry of the Universe is well approximated as spatially homogeneous and isotropic, a so called Friedmann-Lema\^\i tre (FL) Universe~\cite{Fried1,Fried2,Lem1},
\be
ds^2 = a^2(t)\left( -dt^2 +\ga_{ij}dx^idx^j\right) \,.
\ee
Here $t$ is conformal time, $\ga$ is the metric of a 3-space of constant curvature $K$ and $a(t)$ is the scale factor.
For such a highly symmetric spacetime also the energy momentum tensor must obey the same  symmetries and is given by an energy density $\rho(t)=-T_0^0$ and a pressure $P(t)=T^i_i$ (no sum). Einstein's equations become the so called Friedmann equations, where we separate the matter and radiation energy and pressure and the cosmological constant for convenience,
\bea \label{e:F1}
H^2  \equiv \left(\frac{\dot a}{a^2}\right)^2 &=& -\frac{K}{a^2} +\frac{8\pi G}{3}\rho + \frac{\La}{3}\,,\\
\frac{1}{a^2}\frac{d}{dt}\left(\frac{\dot a}{a}\right) &=& -\frac{4\pi G}{3}\left(\rho +3P\right)+ \frac{\La}{3}\,.   \label{e:F2}
\eea
Have we tested these equations in cosmology?

\begin{figure}[!ht]
\begin{center} \includegraphics[width=8cm]{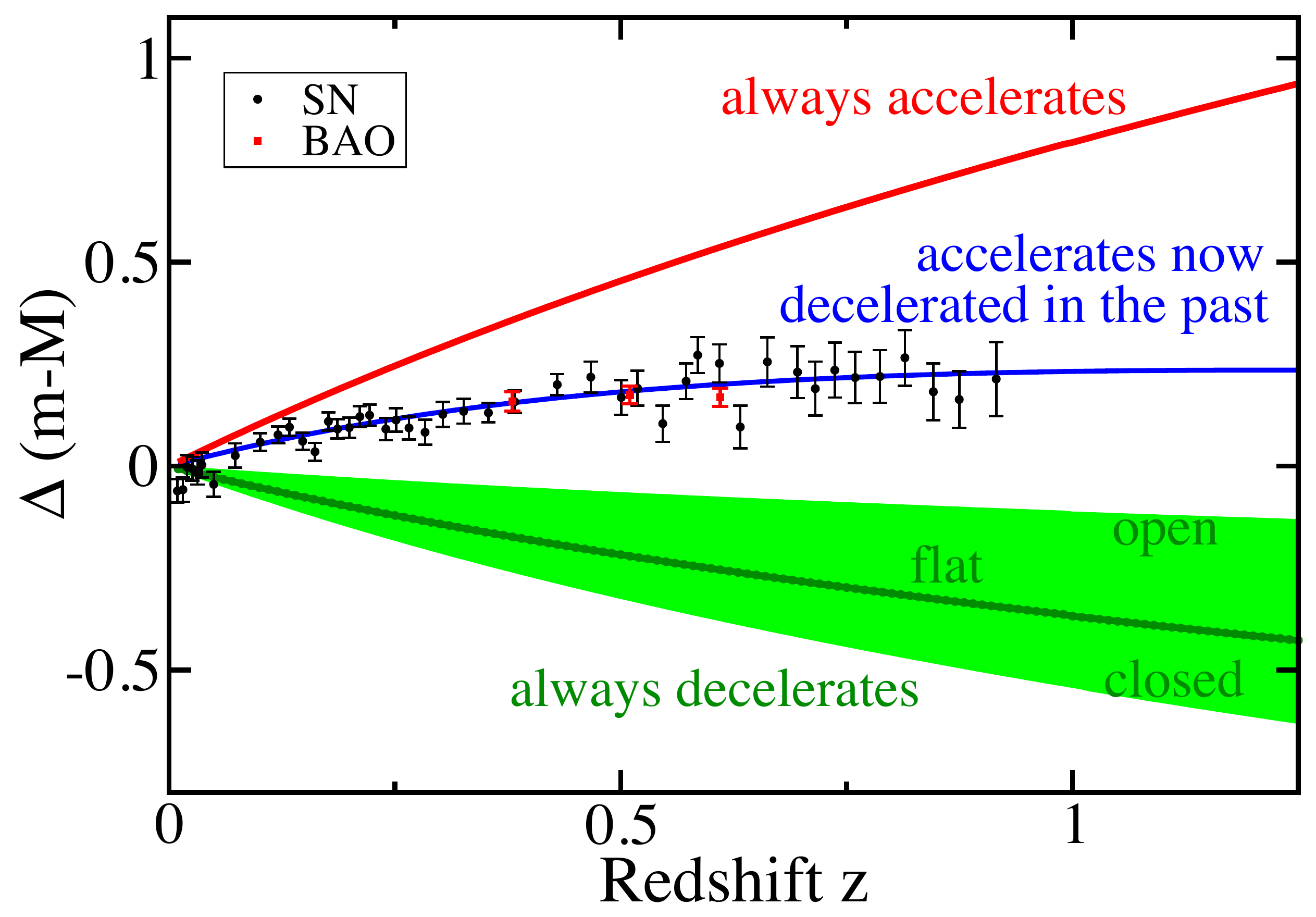}\end{center}
\caption{\label{f:SN} The distance-redshift relation is shown. The  variable plotted in the vertical axis is  $2.5\log_{10}[d_L(z)/d^{\rm Milne}_L(z)]$, where $d^{\rm Milne}_L(z)$ is the luminosity distance in an empty universe which only contains negative curvature, a so called 'Milne universe'. Figure from~\cite{Huterer:2017buf}.}
\end{figure}

We have measured the luminosity distance out to many Cepheids and especially to supernovae  of type Ia up to redshift $z\sim 2.2$, see~\cite{Brout:2022vxf} for a recent analysis. In a FL Universe, the luminosity distance out to redshift $z$ is given by
\be
d_L(z) = (1+z)\chi_K\left(\int_0^z \frac{dz}{H(z)}\right) \,,
\ee
where 
\be
\chi_K(r) = \left\{\begin{array}{cl}
r  & \mbox{if } K=0\\
\frac{1}{\sqrt{K}}\sin(\sqrt{K}r) & \mbox{if } K>0\\
\frac{1}{\sqrt{-K}}\sinh(\sqrt{-K}r)  & \mbox{if } K<0 \,.
\end{array}\right.
\ee
Normalizing the scale factor to $1$ today we have $z+1=1/a$ and measuring $d_L(z)$ precisely for many redshifts $z$, allows us in principle to determine $K$ and $H(z)$.
In order to test Eq.~\eqref{e:F1}, we need an independent measurement of $\rho$ and $\La$. But there is the crux of observational cosmology: If we simply count galaxies and clusters in a large volume and  and assign them a mass which we  roughly infer from their content of stars and gas, we obtain a matter density $\rho$ which is much too small to fit the observed distance $d_L(z)$.  Furthermore, we have absolutely no handle on the cosmological constant. 
Observers actually proceed in the opposite way. They assume that GR is valid and set
\be
H(z) = H_0\sqrt{\Om_K(1+z)^2 + \Om_m(1+z)^3 +\Om_\La}
\ee
where
\be  \Om_K = \frac{-K}{H_0^2}\,, \qquad \Om_m = \frac{8\pi G \rho_0}{3H_0^2}\,, \qquad
 \Om_\La = \frac{\La}{3H_0^2} \,,
\ee
where $\rho_0$ is the present matter density. The first Friedmann equation requires $\Om_K+\Om_m+\Om_\La=1$. Observers
 then fit the three free parameters $\Om_K,~\Om_m$ and $H_0$ to the data.
A reasonably good fit to present data can  be achieved with $\Om_K=0$, $\Om_m=0.3$ and $\Om_\La=0.7$, see~\cite{Brout:2022vxf} for the latest analysis with more than 1500 supernovae Type Ia.  A nice (but not the most recent) fit is shown in Fig.~\ref{f:SN}. There, the observed luminosity distance is compared to the one of an empty universe given by negative curvature only, i.e. with $\Om_K=1$ and $\Om_m=\Om_\La=0$. This is actually simply a part of Minkowski spacetime in accelerated coordinates. It is easy to check that its Riemann tensor vanishes. It is  called the Milne universe. Even if it is not overwhelming, the fit is certainly ok and the data cannot be fitted with $\Om_\La=0$ (the green range in the figure).

Cepheids and Supernovae are so called (modified) standard candles, i.e. objects of which we know the intrinsic luminosity  (or we can infer it from other observables). In cosmology we also have 'standard rulers', i.e. objects of which we know the size. The most accurate is the 'sound horizon', $\la_s(z)$, i.e., the distance a sound wave in the primordial photon-electron-baryon plasma can travel 
from the big bang up to redshift $z$.  This scale manifests itself in the acoustic oscillations of the fluctuation spectrum of the cosmic microwave background (CMB) but also in the galaxy 
correlation function. A standard ruler of size $L$ can be used to determine the angular diameter distance out to redshift $z$. If we see our ruler at redshift $z$ under an angle $\al$ in the sky,  the angular diameter 
distance to it is defined by $d_A(z) = L/\al$. It is related to the luminosity distance via $d_A=d_L/(1+z)^2$. The three red points in Fig.~\ref{f:SN} actually are from the angular diameter distance out to the corresponding redshift as inferred from the acoustic peak in the galaxy correlation 
function. A much more precise value can be inferred from the CMB which yields $d_A(z_{\rm CMB})$ with $z_{\rm CMB} \simeq 1080$. All data together lead to the cosmological parameters~\cite{Planck:2018vyg}
\bea
\Om_K = 0.0007\pm 0.0019\,, \quad \Om_m = 0.348\pm 0.03\,, \quad \Om_\La=1-\Om_K-\Om_m \,.
\eea

Clearly, this data does not test the validity of GR. It assumes GR and infers the matter content of the Universe by fitting the distance redshift relation to a FL Universe with matter and a cosmological constant. The surprising result is that the present expansion rate of the Universe, $H(z)$ for $z<0.5$, is dominated by a cosmological constant or more generally by a substance with  strong negative pressure, $P<-\rho/3$. We call this substance 'dark energy'. 
 Admitting an arbitrary equation of state, $P=w\rho$, the data~\cite{Planck:2018vyg} require  $w=-1.04\pm 0.1$. This is  close to the equation of state of a cosmological constant which has $w=-1$.

So far we have used standard candles and standard rulers to measure the geometry of the background Universe, and we have used Einstein's equations  to infer the matter content of the Universe. Doing so, we have found that the Universe is presently dominated by a component of 
'dark energy' which might be a cosmological constant and the matter content is dominated by 'dark matter' which cannot be the ordinary matter which makes up the stars and gaz in galaxies and clusters but must be something else which interacts much less with photons and ordinary matter and is relatively cold, we termed it cold dark matter (CDM). This has led to a so called cosmological standard model $\La$CDM with the present content of the Universe given by  $\Om_\La \sim 0.68$ $\Om_m\simeq 0.32$ of which the baryons contribute\footnote{The baryon density is 'measured' in the CMB fluctuations with very high precision. Another way to infer it is primordial nucleosynthesis, i.e. the process by which Helium and Deuterium are formed in the early Universe. Both methods agree well. Counting the baryons in stars and hot gaz which make up the galaxies and clusters of galaxies one obtains a smaller value. There are therefore also 'dark baryons' probably in the diffuse  intergalactic medium where hydrogen is mainly ionized. } only $\Om_b \simeq 0.05$ and the rest is CDM. 

Hence only 5\% of the content of the Universe consists of a substance that we are familiar with and that we have seen also via interactions  other  than gravity.  The other 95\%, dark matter and dark energy are inferred by assuming gravity to be governed by the equations of GR. Dark matter is also seen in the velocity dispersion of galaxy clusters as well as in rotation curves of galaxies and even dwarf galaxies which are determined by Newtonian gravity only, but dark energy is inferred solely by cosmological distance measurements and by assuming the Friedmann equations to hold.

It is therefore quite natural to ask  might it be that on large cosmological scales GR is no longer valid? That instead of a cosmological constant we actually see a modification of general relativity?
In the next section we propose that this question can be addressed when studying not only the homogeneous and isotropic background but also its fluctuations.

\section{Large Scale Structure Observations}\label{sec:LSS}
Even if at very large scales, the Universe is close to homogeneous and isotropic, this is not the 
case on small and intermediate scales. There are galaxies, which are very high local over densities of matter and which are often arranged in groups or clusters connected by filaments 
which surround large voids. The hypothesis is that the cosmological large scale structure of the matter distribution (LSS) grew out of small inhomogeneities by gravitational instability. This hypothesis is supported by the 
fact that the fluctuations of the  CMB which represent fluctuations at $z\simeq 1080$ are very small, of order $10^{-5}$. Furthermore, they have precisely the spectrum predicted by a very early 
inflationary phase which amplifies quantum fluctuations and stretches them  into classical fluctuations with a nearly scale invariant spectrum of scalar fluctuations as it is seen in the CMB.

At early times and on large enough scales we can therefore use linear cosmological perturbation theory to study the evolution of fluctuations. At late times, especially on small scales, perturbations grow large and must be studied with numerical N-body simulations. In this paper I shall not discuss this important topic, but let me just mention that after a long and exhaustive investigation of Newtonian N-body simulation, see e.g.~\cite{1985ApJS...57..241E,Angulo:2009rc}, also relativistic N-body simulations are presently under study~\cite{Adamek:2015eda}. These can be used to calculate the observables, which we discuss below within linear perturbations theory, also on smaller, non-linear scales, see~\cite{Lepori:2021lck,Lepori:2020ifz} for first examples. 

Relativistic fluctuation variables are usually gauge dependent, i.e., they depend on the coordinate system used to describe e.g. the background universe. But of course observations are independent of coordinates, i.e., gauge invariant. Therefore it is always possible to describe observables in terms of gauge-invariant quantities.
 
For simplicity, we restrict ourselves to the spatially flat case in this section. The case of non-vanishing curvature is also discussed in the literature, see e.g., Ref.~\cite{DiDio:2016ykq} for the number counts.

\subsection{Number counts}
Let us consider a survey observing galaxies and measuring their directions and redshifts.
A possibility to quantify the  fluctuations in the matter distribution is to count
 the number of galaxies seen in a small solid angle $d\Om$ around a direction $\bn$ in the sky and in a 
 small redshift bin around a redshift $z$. This number is of the form $\bar N(1+\De(\bn,z))d\Om dz$. Here $\bar N(z) \propto \bar\rho(z) r^2(z)H(z)$ is the mean number of galaxies per solid angle 
 and redshift bin. $H(z)$ is the Hubble parameter at redshift $z$ and $r(z)$ is the conformal 
 distance out to redshift $z$, while $\bar\rho$ is the background matter density. The proportionality factor is some mean galaxy mass. In practice, $\bar N$ is directly measured in the survey.
 A detailed study~\cite{Yoo:2009au,Yoo:2010ni,Bonvin:2011bg,Challinor:2011bk,Jeong:2011as} has shown that to first order in perturbation theory, $\De(\bn,z)$ from scalar fluctuations is given by
\bea
\De(\bn,z) &=& bD -3\HH V- (2-5s)\Phi + \Psi + \frac{1}{\HH}
\left[\dot\Phi+ \dd_r(\bV\cdot\bn)\right]  \qquad\qquad \nonumber \\ 
 && %\hspace{-1cm}  
 + \left(\frac{\dot{\HH}}{\HH^2}+\frac{2-5s}{r(z)\HH} +5s-f_{\rm evo}\right)\left(\Psi+ 
\bV\cdot\bn+ 
\int_0^{r(z)}\hspace{-3mm}dr(\dot{\Phi}+\dot{\Psi})\right) 
  \nonumber \\  && %\hspace{-1cm} 
  -\frac{2-5s}{2r(z)}\int_0^{r(z)}\hspace{-2mm}dr \left[\frac{r(z)-r}{r}\Delta_\Om(\Phi+\Psi) -2(\Phi+\Psi)\right]  .    \label{e:Delta}
\eea
Here $D=\de\rho/\bar\rho$ is the density fluctuation in comoving gauge, $b=b(z)$ is the galaxy bias , $V$ is a potential for the peculiar velocity in so called Newtonian or longitudinal gauge, $\bV=-\bnabla V$, $\Phi$ and $\Psi$ are the Bardeen potentials. $\HH(z)= (1+z)H(z)$ is the comoving Hubble parameter,  $s(z)$ is the magnification bias and  $f_{\rm evo}(z)$ is the evolution bias. $\Delta_\Om$ denotes the angular Laplacian. The galaxy bias comes from the fact that fluctuations in the galaxy number density might not be equal to fluctuations in the matter density. There are however good arguments that at scales where linear perturbation theory is valid, the two should at least be proportional, so that $D_{\rm gal}=b(z)D$. The evolution bias is due to the generation and merger of galaxies which makes them scale differently with redshift than the mean density, $\bar\rho$, that scales as $(1+z)^3$. It is defined as
\be
f_{\rm evo}(z) \equiv \HH^{-1}\frac{d\ln(a^3\bar N) }{ dt}  = -(1+z)\frac{d}{dz}\ln\left(\frac{\bar N}{(1+z)^3}\right)  \,.
\ee
The magnification bias $s(z)$ comes from the fact that a galaxy survey is usually flux limited. If  a galaxy is sufficiently amplified by lensing due to  foreground matter, it may make it into our survey even it its unlensed flux would be below the limit of the survey. This quantity depends on the flux limit of the survey considered and on the logarithmic derivative of the mean galaxy number density at this limit. If we see all the galaxies (of a given type), $s(z)=0$. Denoting the  limiting apparent magnitude of the survey by $m_*$, the magnification bias is given by
\be
s(z) = \frac{\dd\log_{10}\bar N(z,m<m_*)}{\dd m_*} \,.
\ee
Note also that in $\De(\bn,z)$ the perturbations variables $D$, $V$, $\Psi$ and $\Phi$ which depend on position $\bx$ and time $t$ have to be evaluated on the background lightcone, $\bx =r(z)\bn$, $t=t(z)=t_0-r(z)$. Also the integrals $dr$ are along the background lightcone, i.e. the integrand is evaluated at $(\bx,t)=(r\bn,t_0-r)$. In \cite{Maartens:2021dqy} the magnification and evolution biases for some near future surveys are estimated. 

The most interesting aspect of Eq.~\eqref{e:Delta} is that it contains not only the density fluctuations $D$ and the velocity field $\bV$ but also the Bardeen potentials $\Phi$ and $\Psi$. In longitudinal gauge these are simply the metric perturbations,
\be
ds^2 = a^2(t)\left[-(1+2\Psi)dt^2+(1-2\Phi)\de_{ij}dx^idx^j\right] \,.
\ee

Hence if we could isolate the different terms, we could in fact measure both, the perturbations of the energy momentum tensor and of the metric.
In the derivation of  Eq.~\eqref{e:Delta} the only assumption on 'matter' that goes in is that galaxies and light follow geodesics. The Einstein equations are never used. Hence \eqref{e:Delta} is valid for any metric theory of gravity. 

The burning question is now~: does the measurement of $\De(\bn, z)$ for many different redshifts and directions allow us to isolate its different contributions from $D$, $V$, $\Phi$ and $\Psi$?
The short answer is NO.  But let us study in somewhat more detail what we can do.

First we need to stress that our theories of the generation and evolution of cosmological perturbations do not predict the values of any perturbation variable at given positions and times, e.g., 
$\Psi(\bx,t)$, but they only provide statistical ensemble averages like, e.g., power spectra or correlation functions. By definition, the mean of $\De(\bn,z)$ vanishes, but its correlation or power spectra do not.
At fixed redshift, $z$, $\De(\bn,z)$ is a function on the sphere which we can expand is spherical harmonics,
\be
\De(\bn,z) =\sum a_{\ell m}(z)Y_{\ell m}(\bn) \,.
\ee
If we require statistical isotropy, the expectation values of the $a_{\ell m}$'s vanish and
\be
\langle  a_{\ell m}(z) a^*_{\ell' m'}(z')\rangle = \de_{\ell\ell'}\de_{m m'}C_\ell(z,z') \,.
\ee
The $C_\ell(z,z')$'s are the angular power spectra of $\De(\bn,z)$.
The correlation function is then given by
\be
\langle\De(\bn,z)\De(\bn',z')\rangle = \frac{1}{4\pi}\sum_\ell (2\ell+1)C_\ell(z,z') P_\ell(\bn\cdot\bn')\,,
\ee
where $P_\ell$ is the Legendre polynomial of order $\ell$. The fact that $\langle\De(\bn,z)\De(\bn',z')\rangle$ depends on directions only via $\bn\cdot\bn'$ is again a consequence of statistical isotropy.

Of course we can not really measure an ensemble average, but each  measured
$a_{\ell m}(z) a^*_{\ell m}(z')$ provides an  independent estimator of  $C_\ell(z,z')$.
If perturbations are Gaussian, which is a good approximation on linear scales, the power spectra contain all the statistical information. The error due to the fact that there are at best $2\ell+1$ different $a_{\ell m}(z)$'s is $\sqrt{2/(2\ell+1)f_{\rm sky}}$, where $f_{\rm sky}$ denotes the observed sky fraction. This minimal error is called 'cosmic variance'. Hence the low multipoles always have very large errors due to the fact that we have only very few estimators of them in the observable Universe.

Studying the relation of metric perturbations and density perturbations via Einstein's equations one finds that $\Psi\,,\Phi \sim (\HH/k)^2 D$. Hence on subhorizon scales, $k\gg \HH$, metric perturbations are substantially suppressed. Therefore, the terms in Eq.~\eqref{e:Delta} containing the gravitational potential are typically relevant only on very large scales, i.e. very low $\ell$'s for which cosmic variance is large. There is one exception to this which is the lensing term given by the Newtonian convergence\footnote{I call this the Newtonian convergence since the full convergence which appears in the Jacobi map is given by all the terms which multiply the factor $-5s$, see, e.g.~\cite{Fanizza:2022wob}. In a quasi-Newtonian study of lensing, however, one only obtains $\ka^{(N)}$, see, e.g.,~\cite{Bartelmann:2010fz}. } $\ka^{(N)}$,
\be
\frac{2-5s}{2}\int_0^{r(z)}\hspace{-1mm}dr\frac{r(z)-r}{r(z)r}\Delta_\Om(\Phi+\Psi)
=(2-5s)\ka^{(N)}(\bn,z)\,.
\ee
For a given harmonic $\ell$ the angular Laplacian $\Delta_\Om$ yields a factor $-\ell(\ell+1)$ which is of the order $(kr(z))^2 \sim (k/\HH(z))^2$, for not too small redshifts. However, as it is an integrated term, cancellations happen and reduce this term typically to the order of $(k/\HH(z))\Psi$. What is more interesting, as this is an integrated term, it contributes significantly to $C_\ell(z,z')$ for $z\neq z'$, while the non-integrated terms are very small for well separated redshift bins.
It has been argued that this term will be measured well in future survey, especially in the relatively wide redshift bins of photometric galaxy surveys~\cite{Montanari:2015rga,Jelic-Cizmek:2020pkh,Euclid:2021rez}. It has already been measured in the past by correlating foreground galaxies e.g. with background quasars, see, e.g.,~\cite{Menard:2002da}.

In the auto correlations, $z=z'$, one  mainly measure the Newtonian contributions,
\be\label{e:DV}
\De^{(N)}(\bn,z) = bD + \frac{1}{\HH}\dd_r(\bV\cdot\bn)
\ee
which are well known~\cite{Kaiser1987} and have been measured very precisely in the power spectrum in redshift space.  The density term prominently shows the so called baryon acoustic oscillations (BAO) which are a remnant from the coupled baryon photon plasma which performed acoustic oscillations  before decoupling.  Since the acoustic horizon is well known, the angular scale of these oscillations  yields an excellent measure of the angular diameter distance, $d_A(z)$.  The three red points in Fig.~\ref{f:SN} are from such measurements with the BOSS \cite{BOSS:2016wmc} survey. In the future it will also be possible to measure the radial redshift difference corresponding to this scale,
\be
\la_s(z) = r(z+\De z/2)-  r(z-\De z/2) \simeq \De z/H(z)\,.
\ee
 This will provide a measure of the Hubble parameter at redshift $z$.

The velocity term in \eqref{e:DV} is  called the redshift space distortion (RSD). It comes from the fact that galaxies with radial peculiar velocities are seen at slightly higher, respectively lower redshift. This leads to a radial volume distortion, the RSD. It can be  measured in the correlation function. When observing a small patch of sky which can be treated within the flat sky approximation, \\  
$\frac{1}{\HH}\dd_r(\bV\cdot\bn) = \mu^2f(z)D$, where $\mu$ is the cosine of the angle between the observation  direction and the velocity. Here  
$$ f(z) = \frac{d\ln D_1}{d\ln a} \simeq \left(\Om_m(z)\right)^{0.55} $$
is the linear growth rate of density fluctuations, and we have used that within linear perturbation theory the continuity equation relates density fluctuations to the velocity potential via  $V = \HH(z)f(z)D$. This growth rate is very sensitive to the theory of gravity and measuring it is an excellent test of GR.  Present measurements of the growth rate are, however, degenerate with the amplitude of perturbations cast in $\si_8$. Here $\si_8^2$ is the variance of density fluctuations inside a ball of radius 8Mpc.

\begin{figure}
\begin{center}
\includegraphics[width=8cm]{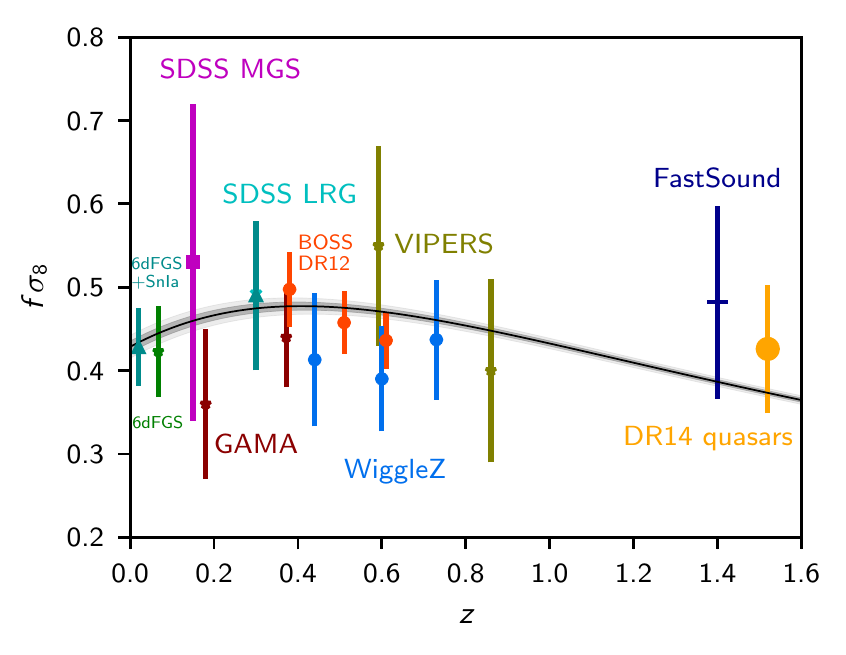}
\end{center}
\caption{Measurements of $f\si_8$ at different redshifts from different surveys. The grey line shows the behaviour of $f(z)\si_8(z)$ for the best fit $\La$CDM model from Planck.  Figure from~\cite{Planck:2018vyg}. \label{f:growth} }
\end{figure}

In a small survey which can be treated with the flat sky approximation, while the quadrupole and hexadecapole in  the correlation function are  excellent measures of the growth rate, while the monopole is used to identify the BAO  feature. Several recent surveys give rather accurate determinations of $\si_8f(z)$, see~\cite{Beutler:2012px,WiggleZ:2013akc,Samushia:2013yga,Reid:2014iaa,BOSS:2016ntk,  Mohammad:2017lzz,Shi:2017qpr,Ruggeri:2018kdn,Bautista:2020ahg,Lange:2021zre}.
Many recent measurements are shown in Fig.~\ref{f:growth}.

With cosmological galaxy number counts we can therefore measure density fluctuations, the growth rate or, more generically, velocity perturbations and the lensing potential. Knowing the lensing potential out to many different redshifts we can use these measurements to infer the so called Weyl potential,
\be
\Phi_W = \frac{1}{2}(\Phi+\Psi) \,.
\ee
In $\La$CDM cosmology, where anisotropic stresses (nearly) vanish at late time, 
\be\label{e:poteq}
\Phi\simeq\Psi\simeq\Phi_W\,,
\ee 
with relative differences of the order of 0.1\%.
Furthermore, within a metric theory of gravity, where galaxies move on geodesics,  $d(a\bV)/dt = -a\bnabla\Psi$. Hence measuring the velocity via redshift space distortions at many different redshifts, allows in principle also to isolate $\Psi$ and so to test  \eqref{e:poteq}. 
A way to measure the velocity via the 'Doppler term', $\bV\cdot\bn$ in Eq.~\eqref{e:Delta} has been suggested in~\cite{Bonvin:2013ogt,Gaztanaga:2015jrs}. Since this term is odd in $\bn$,
it can be isolated as the antisymmetric part in the correlation function of two different tracers, e.g. red and blue or luminous and faint galaxies. A violation of \eqref{e:poteq} could either indicate a violations of the equivalence principle for dark matter, see also~\cite{Bonvin:2018ckp}, or a modification of the theory of gravity which usually implies $\Phi\neq\Psi$.
Together with the growth rate this already provides two excellent tests of general relativity.

Very recently, the idea to weigh cosmological number counts, e.g., with the redshift has been explored~\cite{Legrand:2020sek,Matthewson:2022bpp}, and it has been show that this can lead to interesting complementary cosmological information. It will be important to investigate these new observables in more detail, also in view of the their power to test GR.  

\subsection{Intensity mapping}
Mapping out individual galaxies in a survey and measuring their redshift is very costly.
A 'cheaper' alternative to galaxy number counts is the so called intensity mapping.
In this technique we do not count galaxies, but just measure the intensity of the sky brightness in a specific, strong atomic or molecular line, e.g., the 21cm line   from the hyperfine spin-flip transition of neutral hydrogen~\cite{Chang:2007xk,Wyithe:2008mv}. This yields the surface density of hydrogen in the sky at a fixed redshift. The difficulty here is the foreground removal, as usually the foreground is many orders of magnitude larger than the signal.  Foreground removal can be achieved by requesting a very strong intensity change as a function of wavelength as it is only present in lines.

At  high redshift, before reionization, $z>6.5$, intensity mapping of the 21cm line is interesting by itself as it can measure the evolution of small, linear density fluctuations. Around $z\sim 6$ one can investigate the process of reionization by this method. After reionization, $z<6$, neutral hydrogen is mainly present in galaxies and proto-galaxies, and 21cm intensity mapping is actually a coarse grained mapping of large scale structure, see~\cite{Furlanetto:2006jb} for an overview.

So far, most  measurements have correlated 21cm emission with existing galaxy surveys~\cite{10.1111/j.1745-3933.2008.00581.x,Masui_2013,Tramonte:2020csa}. A very surprising result~\cite{Bowman:2018yin} is a  measurement of a 21cm signal at $z\simeq 17$ which has more than twice the expected intensity. This result has, however, not been confirmed (nor refuted) until now. Several experiments to directly detect large scale structure with intensity mapping are presently being prepared and shall take data in the near future~\cite{Kuhn:2021vgx,Yohana:2019ahg,Barry:2021szi}. A first result has very recently been put on the arXiv~\cite{CHIME:2022kvg}. Forecasts for the prospects of intensity mappings 
are detailed in several papers-\cite{Santos:2015gra,Abdalla:2015zra,Fonseca:2018hsu}

The HI intensity is usually cast into an equivalent temperature $T_{\rm HI}$ and its fluctuation is nearly proportional to the fluctuation in the number of galaxies. The main difference is that intensity 
is determined per surface area and, as is well known, gravitational lensing conserves surface brightness. Therefore, at first order in perturbations theory, like for the CMB, there are no lensing terms in intensity mapping.
The relative fluctuations in intensity are therefore given exactly by Eq.~\eqref{e:Delta} when setting the magnifications bias $s\equiv 2/5$. This result is derived in detail in \cite{Hall:2012wd}.
Therefore, while intensity mapping cannot be used to determine the convergence $\ka$, it will be very useful for density and RSD. Of course also the bias will be the one of hydrogen, $b_{\rm HI}$ which is usually somewhat different from the galaxy bias $b$.

Also the cross-correlations of intensity maps and galaxy number counts will prove very useful for
velocity measurements~\cite{Hall:2016bmm} and for identifying $\ka^{(N)}$ in the number count surveys~\cite{Jalilvand:2019bhk}.

Of course also other lines, not only the 21cm line of hydrogen, can be used for intensity mapping, see~\cite{Kovetz:2019uss} for an overview.

\subsection{Shear Measurements}
The foreground gravitational field acts like a medium with spatially dependent index of refraction on light rays which come to us from far away sources. The images are stretches, sheared and can also be rotated. For small images this is described with the so called Jacobi map~\cite{Perlick:2004tq}.
To first order in cosmological perturbation theory there is no rotation~\cite{Fanizza:2022wob}. The shear from scalar perturbations is determined fully by the lensing potential
\be
\phi = \int_0^{r(z)}\hspace{-1mm}dr\frac{r(z)-r}{r(z)r}(\Phi+\Psi) \,.
\ee
The shear is usually cast as a complex helicity-2 variable defined as
\be
\ga = (\nabla_1^2-\nabla_2^2+i\nabla_1\nabla_2)\phi \,,
\ee
where $\nabla_i$ is the (covariant) derivative in direction $\bfe_i$ and $(\bfe_1,\bfe_2)$ form an orthonormal basis on the sphere.
Even though $\ga$ depends on our choice of basis, once we determine the correlation function $\langle\ga(\bn,z)\ga(\bn',z')\rangle$, we can choose $\bfe_1$ as the vector pointing from the point $\bn$ to $\bn'$ on the sphere (more precisely as the tangent vector to the geodesic from $\bn$ to $\bn'$), which defines the basis intrinsically.

The ellipticity of background galaxies is affected by the foreground shear, see~\cite{Bartelmann:1999yn,Bartelmann:2010fz,Kilbinger:2014cea} for reviews on the subject. Several recent measurements of the shear correlation function and the shear angular power spectrum in different redshift bins have been performed. Some recent results are found in Refs.~\cite{Hildebrandt:2018yau,KiDS:2020suj,DES:2020daw,DES:2021wwk,Amon:2022ycy,DES:2022qpf}. These have usually been analysed with the assumptions of purely scalar perturbations in $\La$CDM, or one of its variants like $w$CDM, where the cosmological constant is replaced by a dark energy component with the equation of state $P=w\rho$. In these studies the authors have set (in Fourier space) $k^2\Phi=k^2\Psi = -(3/2)(H_0/k)^2\Om_m (1+z)D$ and used their results to constrain a variable closely related to $\Om_mD$, more precisely they measured
\be
S_8 = \si_8\Om_m^{1/2} \,.
\ee
Their results are  somewhat in tension with the value of $S_8$ inferred from Planck.  They obtain a best fit value for $S_8$ that is typically about 2 standard deviation below the Planck result.

However, before jumping to conclusions, it must be noted that cosmic shear measurements are very difficult. They have been proposed for the first time in the 60ties~\cite{Kristian:1965sz} and first serious attempts to measure shear were made in the 80ties, see, e.g.,~\cite{1983ApJ...271..431V}. However,  first measurements of the shear correlation function have come out only a couple years ago. The reason for this is mainly intrinsic alignement: If the ellipticity of two nearby galaxies is correlated this can have two reasons. Either the fact that we see them through the same foreground shear field or that they have been formed from the same large scale over density and are intrinsically aligned. Present shear analysis is always combined with a model for intrinsic alignement which has several free parameters that have to be  fitted simultaneously with the cosmological parameters. Nevertheless, it is not clear whether the intrinsic alignement model is sufficiently detailed to capture all the relevant physics, or whether it may lead to a systematic error in our estimate of $S_8$.

Recently, a new method to measure cosmic shear via its rotation of the principal axis of a galaxy has been proposed~\cite{Francfort:2022laa}. For this, however, one needs another direction which keeps the information of the original orientation of the galaxy in the source plane. This can be provided by the polarization of light which is parallel transported along the light ray. The polarization direction, e.g., of radio galaxies is correlated with their semi-minor axis. Simultaneous determination of the semi-minor axis in the image and of the polarization direction  can then be used to determine the rotation of this axis due to shear. This measure is not plagued by intrinsic alignement. It will have to be investigated in detail what its systematic uncertainty are and what its potential is to improve shear measurements.

If the shear is measured it can be used as an independent measurement of the Weyl potential, but it can also be used to test its relation with $\ka^{(N)}$ measured on galaxy number counts.  For purely scalar perturbations their power spectra should be related as, see e.g.~\cite{Stebbins:1996wx}
\be\label{e:kaga}
C_\ell^{\ka^{(N)}}(z,z') = \frac{\ell(\ell+1)}{(\ell+2)(\ell-1)}C_\ell^\ga \,.
\ee
It has been shown, see e.g.,~\cite{Fanizza:2022wob} that at very large scales, low $\ell$, $\ell\simeq 20$ this relation does not hold for the full convergence in the Jacobi map, however at higher values of $\ell$ it is very accurate.

Testing this relation will be very important. If it does not hold this means that either vector or tensor fluctuations are relevant for which this relation is not valid, see~\cite{Schmidt:2012ne,Fanizza:2022wob} or then that photons do not move on geodesics but are subject to some 'fifth force' or similar.  The second would be a very significant deviation from GR, where not Einstein's equations are put in question, these are not used for relation \eqref{e:kaga}, but the 'equivalence principle' for photons. But also the first would be interesting and might hint, e.g., to a stochastic gravitational wave background.

\section{Conclusion}\label{sec:con}
%{\tt putting it all together we can measure $D$, $V$ and $\Phi$ etc...}
In this paper we have shown with several examples that measurements of cosmological perturbations or more precisely their correlation functions and power spectra can be used to measure both, components of the matter energy momentum tensor and components of the metric. In this way they open the route to test General Relativity  on cosmological scales. We have only mentioned the best known examples here but there are several more. Furthermore, appart from the two point correlation function and the power spectrum, there are other observables. For example the higher order statistical quantities, $N$-point correlation functions  and the corresponding $N$-point spectra as well as other expectation values for which we can develop estimators, like,  e.g.,  Minkowski functionals, see~\cite{Pranav:2018pnu,Liu:2022vtr}, or wavelets. $N$-point correlation functions are especially relevant in the non-linear regime where deviations from Gaussian statistics become very significant and  contain independent information. In particular the three-point function or the corresponding power spectrum, the bi-spectrum, see e.g.~\cite{DiDio:2018unb,deWeerd:2019cae,Durrer:2020orn}, have very specific properties for example in the squeezed limit where the wavelength of one density fluctuation is much larger than the other two. These properties are a simple consequence of the equivalence principle and can therefore be used to test it.
Another interesting observable is the correlation between CMB anisotropies and galaxy number counts which are due to lensing of the CMB by foreground galaxies, see e.g.~\cite{Bermejo-Climent:2021jxf}.
\vspace{0.2cm}

In the past decades we used cosmological observations mainly to determine cosmological parameters with good precision. The excellent CMB data has allowed us to come up with a simple cosmological standard model, flat $\La$CDM, that fits the CMB data very well and requires only 6 parameters. Even though discrepancies of local measurements of the Hubble constant~\cite{Riess:2021jrx} might hint to shortcomings of the model, there is also 'the elephant in the room':  The $\La$CDM model contains about 95\% 'dark content', dark energy  in the form of  
the cosmological constant $\La$ and cold dark matter, which have not been detected in any other way than via their gravitational interaction; this is certainly unsatisfactory. It is mainly this fact that motivates us to use present and future cosmological observations not only to improve on the precision of the cosmological model parameters, but much more to test the  basic underpinning of $\La$CDM, Einstein's theory of General Relativity.
\vspace*{0.3cm}

\backmatter

\bmhead{Acknowledgments}
Special thanks go to Somak Raychaudhury and Kandaswamy Subramanian from IUCAA who invited me to give them a colloquium on this topic, on which the present article is based.
This work is supported by the Swiss National Science Foundation.

%\begin{appendices}

%\section{Section title of first appendix}\label{secA1}

%An appendix contains supplementary information that is not an essential part of the text itself but which may be helpful in providing a more comprehensive understanding of the research problem or it is information that is too cumbersome to be included in the body of the paper.




%\end{appendices}

\bibliography{sn-bibliography}% common bib file


\begin{thebibliography}{84}
\ifx \bisbn   \undefined \def \bisbn  #1{ISBN #1}\fi
\ifx \binits  \undefined \def \binits#1{#1}\fi
\ifx \bauthor  \undefined \def \bauthor#1{#1}\fi
\ifx \batitle  \undefined \def \batitle#1{#1}\fi
\ifx \bjtitle  \undefined \def \bjtitle#1{#1}\fi
\ifx \bvolume  \undefined \def \bvolume#1{\textbf{#1}}\fi
\ifx \byear  \undefined \def \byear#1{#1}\fi
\ifx \bissue  \undefined \def \bissue#1{#1}\fi
\ifx \bfpage  \undefined \def \bfpage#1{#1}\fi
\ifx \blpage  \undefined \def \blpage #1{#1}\fi
\ifx \burl  \undefined \def \burl#1{\textsf{#1}}\fi
\ifx \doiurl  \undefined \def \doiurl#1{\url{https://doi.org/#1}}\fi
\ifx \betal  \undefined \def \betal{\textit{et al.}}\fi
\ifx \binstitute  \undefined \def \binstitute#1{#1}\fi
\ifx \binstitutionaled  \undefined \def \binstitutionaled#1{#1}\fi
\ifx \bctitle  \undefined \def \bctitle#1{#1}\fi
\ifx \beditor  \undefined \def \beditor#1{#1}\fi
\ifx \bpublisher  \undefined \def \bpublisher#1{#1}\fi
\ifx \bbtitle  \undefined \def \bbtitle#1{#1}\fi
\ifx \bedition  \undefined \def \bedition#1{#1}\fi
\ifx \bseriesno  \undefined \def \bseriesno#1{#1}\fi
\ifx \blocation  \undefined \def \blocation#1{#1}\fi
\ifx \bsertitle  \undefined \def \bsertitle#1{#1}\fi
\ifx \bsnm \undefined \def \bsnm#1{#1}\fi
\ifx \bsuffix \undefined \def \bsuffix#1{#1}\fi
\ifx \bparticle \undefined \def \bparticle#1{#1}\fi
\ifx \barticle \undefined \def \barticle#1{#1}\fi
\bibcommenthead
\ifx \bconfdate \undefined \def \bconfdate #1{#1}\fi
\ifx \botherref \undefined \def \botherref #1{#1}\fi
\ifx \url \undefined \def \url#1{\textsf{#1}}\fi
\ifx \bchapter \undefined \def \bchapter#1{#1}\fi
\ifx \bbook \undefined \def \bbook#1{#1}\fi
\ifx \bcomment \undefined \def \bcomment#1{#1}\fi
\ifx \oauthor \undefined \def \oauthor#1{#1}\fi
\ifx \citeauthoryear \undefined \def \citeauthoryear#1{#1}\fi
\ifx \endbibitem  \undefined \def \endbibitem {}\fi
\ifx \bconflocation  \undefined \def \bconflocation#1{#1}\fi
\ifx \arxivurl  \undefined \def \arxivurl#1{\textsf{#1}}\fi
\csname PreBibitemsHook\endcsname

%%% 1
\bibitem{astro1}
\begin{bbook}
\bauthor{\bsnm{{Padmanabhan, T. }}}:
\bbtitle{Theoretical Astrophysics Volume I: Astrophysical Processes}.
\bpublisher{Cambridge University Press},
\blocation{Cambridge}
(\byear{2000})
\end{bbook}
\endbibitem

%%% 2
\bibitem{astro2}
\begin{bbook}
\bauthor{\bsnm{{Padmanabhan, T. }}}:
\bbtitle{Theoretical Astrophysics Volume II: Stars Ans Stellar Systems}.
\bpublisher{Cambridge University Press},
\blocation{Cambridge}
(\byear{2001})
\end{bbook}
\endbibitem

%%% 3
\bibitem{astro3}
\begin{bbook}
\bauthor{\bsnm{{Padmanabhan, T. }}}:
\bbtitle{Theoretical Astrophysics Volume III: Galaxies and Cosmology}.
\bpublisher{Cambridge University Press},
\blocation{Cambridge}
(\byear{2002})
\end{bbook}
\endbibitem

%%% 4
\bibitem{rela}
\begin{bbook}
\bauthor{\bsnm{{Padmanabhan, T. }}}:
\bbtitle{Gravitation: Foundations and Frontiers}.
\bpublisher{Cambridge University Press},
\blocation{Cambridge}
(\byear{2010})
\end{bbook}
\endbibitem

%%% 5
\bibitem{Fried1}
\begin{barticle}
\bauthor{\bsnm{{Friedmann, A. }}}:
\batitle{{\"Uber die Kr\"umming des Raumes}}.
\bjtitle{Z. Phys.}
\bvolume{10},
\bfpage{377}
(\byear{1922})
\end{barticle}
\endbibitem

%%% 6
\bibitem{Fried2}
\begin{barticle}
\bauthor{\bsnm{{Friedmann, A. }}}:
\batitle{{\"Uber die M\"oglichkeit einer Welt mit konstanter negativer
  Kr\"umming des Raumes}}.
\bjtitle{Z. Phys.}
\bvolume{21},
\bfpage{326}
(\byear{1924})
\end{barticle}
\endbibitem

%%% 7
\bibitem{Lem1}
\begin{barticle}
\bauthor{\bsnm{{Lema\^\i tre, G. }}}:
\batitle{L'univers en expansion}.
\bjtitle{Ann. Soc. Bruxelles}
\bvolume{47A},
\bfpage{49}
(\byear{1927})
\end{barticle}
\endbibitem

%%% 8
\bibitem{Brout:2022vxf}
\begin{botherref}
\oauthor{\bsnm{{Brout, Dillon and others }}}:
{The Pantheon+ Analysis: Cosmological Constraints}
(2022)
{\href{https://arxiv.org/abs/2202.04077}{{arXiv:2202.04077}}}
{[astro-ph.CO]}
\end{botherref}
\endbibitem

%%% 9
\bibitem{Huterer:2017buf}
\begin{barticle}
\bauthor{\bsnm{{Huterer, Dragan and Shafer, Daniel L }}}:
\batitle{{Dark energy two decades after: Observables, probes, consistency
  tests}}.
\bjtitle{Rept. Prog. Phys.}
\bvolume{81}(\bissue{1}),
\bfpage{016901}
(\byear{2018})
{\href{https://arxiv.org/abs/1709.01091}{{arXiv:1709.01091}}}
{[astro-ph.CO]}.
\doiurl{10.1088/1361-6633/aa997e}
\end{barticle}
\endbibitem

%%% 10
\bibitem{Planck:2018vyg}
\begin{barticle}
\bauthor{\bsnm{Aghanim}, \binits{N.}}, \betal:
\batitle{{Planck 2018 results. VI. Cosmological parameters}}.
\bjtitle{Astron. Astrophys.}
\bvolume{641},
\bfpage{6}
(\byear{2020})
{\href{https://arxiv.org/abs/1807.06209}{{arXiv:1807.06209}}}
{[astro-ph.CO]}.
\doiurl{10.1051/0004-6361/201833910}.
\bcomment{[Erratum: Astron.Astrophys. 652, C4 (2021)]}
\end{barticle}
\endbibitem

%%% 11
\bibitem{1985ApJS...57..241E}
\begin{barticle}
\bauthor{\bsnm{{Efstathiou}}, \binits{G.}},
\bauthor{\bsnm{{Davis}}, \binits{M.}},
\bauthor{\bsnm{{White}}, \binits{S.D.M.}},
\bauthor{\bsnm{{Frenk}}, \binits{C.S.}}:
\batitle{{Numerical techniques for large cosmological N-body simulations}}.
\bjtitle{Astrophys. J.}
\bvolume{57},
\bfpage{241}--\blpage{260}
(\byear{1985}).
\doiurl{10.1086/191003}
\end{barticle}
\endbibitem

%%% 12
\bibitem{Angulo:2009rc}
\begin{barticle}
\bauthor{\bsnm{Angulo}, \binits{R.E.}},
\bauthor{\bsnm{White}, \binits{S.D.M.}}:
\batitle{{One simulation to fit them all - changing the background parameters
  of a cosmological N-body simulation}}.
\bjtitle{Mon. Not. Roy. Astron. Soc.}
\bvolume{405},
\bfpage{143}
(\byear{2010})
{\href{https://arxiv.org/abs/0912.4277}{{arXiv:0912.4277}}}
{[astro-ph.CO]}.
\doiurl{10.1111/j.1365-2966.2010.16459.x}
\end{barticle}
\endbibitem

%%% 13
\bibitem{Adamek:2015eda}
\begin{barticle}
\bauthor{\bsnm{Adamek}, \binits{J.}},
\bauthor{\bsnm{Daverio}, \binits{D.}},
\bauthor{\bsnm{Durrer}, \binits{R.}},
\bauthor{\bsnm{Kunz}, \binits{M.}}:
\batitle{{General relativity and cosmic structure formation}}.
\bjtitle{Nature Phys.}
\bvolume{12},
\bfpage{346}--\blpage{349}
(\byear{2016})
{\href{https://arxiv.org/abs/1509.01699}{{arXiv:1509.01699}}}
{[astro-ph.CO]}.
\doiurl{10.1038/nphys3673}
\end{barticle}
\endbibitem

%%% 14
\bibitem{Lepori:2021lck}
\begin{barticle}
\bauthor{\bsnm{Lepori}, \binits{F.}},
\bauthor{\bsnm{Adamek}, \binits{J.}},
\bauthor{\bsnm{Durrer}, \binits{R.}}:
\batitle{{Cosmological simulations of number counts}}.
\bjtitle{JCAP}
\bvolume{12}(\bissue{12}),
\bfpage{021}
(\byear{2021})
{\href{https://arxiv.org/abs/2106.01347}{{arXiv:2106.01347}}}
{[astro-ph.CO]}.
\doiurl{10.1088/1475-7516/2021/12/021}
\end{barticle}
\endbibitem

%%% 15
\bibitem{Lepori:2020ifz}
\begin{barticle}
\bauthor{\bsnm{Lepori}, \binits{F.}},
\bauthor{\bsnm{Adamek}, \binits{J.}},
\bauthor{\bsnm{Durrer}, \binits{R.}},
\bauthor{\bsnm{Clarkson}, \binits{C.}},
\bauthor{\bsnm{Coates}, \binits{L.}}:
\batitle{{Weak-lensing observables in relativistic N-body simulations}}.
\bjtitle{Mon. Not. Roy. Astron. Soc.}
\bvolume{497}(\bissue{2}),
\bfpage{2078}--\blpage{2095}
(\byear{2020})
{\href{https://arxiv.org/abs/2002.04024}{{arXiv:2002.04024}}}
{[astro-ph.CO]}.
\doiurl{10.1093/mnras/staa2024}
\end{barticle}
\endbibitem

%%% 16
\bibitem{DiDio:2016ykq}
\begin{barticle}
\bauthor{\bsnm{Di~Dio}, \binits{E.}},
\bauthor{\bsnm{Montanari}, \binits{F.}},
\bauthor{\bsnm{Raccanelli}, \binits{A.}},
\bauthor{\bsnm{Durrer}, \binits{R.}},
\bauthor{\bsnm{Kamionkowski}, \binits{M.}},
\bauthor{\bsnm{Lesgourgues}, \binits{J.}}:
\batitle{{Curvature constraints from Large Scale Structure}}.
\bjtitle{JCAP}
\bvolume{06},
\bfpage{013}
(\byear{2016})
{\href{https://arxiv.org/abs/1603.09073}{{arXiv:1603.09073}}}
{[astro-ph.CO]}.
\doiurl{10.1088/1475-7516/2016/06/013}
\end{barticle}
\endbibitem

%%% 17
\bibitem{Yoo:2009au}
\begin{barticle}
\bauthor{\bsnm{Yoo}, \binits{J.}},
\bauthor{\bsnm{Fitzpatrick}, \binits{A.L.}},
\bauthor{\bsnm{Zaldarriaga}, \binits{M.}}:
\batitle{{A New Perspective on Galaxy Clustering as a Cosmological Probe:
  General Relativistic Effects}}.
\bjtitle{Phys.Rev.}
\bvolume{D80},
\bfpage{083514}
(\byear{2009})
{\href{https://arxiv.org/abs/0907.0707}{{arXiv:0907.0707}}}
{[astro-ph.CO]}.
\doiurl{10.1103/PhysRevD.80.083514}
\end{barticle}
\endbibitem

%%% 18
\bibitem{Yoo:2010ni}
\begin{barticle}
\bauthor{\bsnm{Yoo}, \binits{J.}}:
\batitle{{General Relativistic Description of the Observed Galaxy Power
  Spectrum: Do We Understand What We Measure?}}
\bjtitle{Phys.Rev.}
\bvolume{D82},
\bfpage{083508}
(\byear{2010})
{\href{https://arxiv.org/abs/1009.3021}{{arXiv:1009.3021}}}
{[astro-ph.CO]}.
\doiurl{10.1103/PhysRevD.82.083508}
\end{barticle}
\endbibitem

%%% 19
\bibitem{Bonvin:2011bg}
\begin{barticle}
\bauthor{\bsnm{Bonvin}, \binits{C.}},
\bauthor{\bsnm{Durrer}, \binits{R.}}:
\batitle{{What galaxy surveys really measure}}.
\bjtitle{Phys.Rev.}
\bvolume{D84},
\bfpage{063505}
(\byear{2011})
{\href{https://arxiv.org/abs/arXiv:1105.5280}{{arXiv:arXiv:1105.5280}}}
{[astro-ph.CO]}.
\doiurl{10.1103/PhysRevD.84.063505}
\end{barticle}
\endbibitem

%%% 20
\bibitem{Challinor:2011bk}
\begin{barticle}
\bauthor{\bsnm{Challinor}, \binits{A.}},
\bauthor{\bsnm{Lewis}, \binits{A.}}:
\batitle{{The linear power spectrum of observed source number counts}}.
\bjtitle{Phys.Rev.}
\bvolume{D84},
\bfpage{043516}
(\byear{2011})
{\href{https://arxiv.org/abs/arXiv:1105.5292}{{arXiv:arXiv:1105.5292}}}
{[astro-ph.CO]}.
\doiurl{10.1103/PhysRevD.84.043516}
\end{barticle}
\endbibitem

%%% 21
\bibitem{Jeong:2011as}
\begin{barticle}
\bauthor{\bsnm{Jeong}, \binits{D.}},
\bauthor{\bsnm{Schmidt}, \binits{F.}},
\bauthor{\bsnm{Hirata}, \binits{C.M.}}:
\batitle{{Large-scale clustering of galaxies in general relativity}}.
\bjtitle{Phys. Rev. D}
\bvolume{85},
\bfpage{023504}
(\byear{2012})
{\href{https://arxiv.org/abs/1107.5427}{{arXiv:1107.5427}}}
{[astro-ph.CO]}.
\doiurl{10.1103/PhysRevD.85.023504}
\end{barticle}
\endbibitem

%%% 22
\bibitem{Maartens:2021dqy}
\begin{barticle}
\bauthor{\bsnm{Maartens}, \binits{R.}},
\bauthor{\bsnm{Fonseca}, \binits{J.}},
\bauthor{\bsnm{Camera}, \binits{S.}},
\bauthor{\bsnm{Jolicoeur}, \binits{S.}},
\bauthor{\bsnm{Viljoen}, \binits{J.-A.}},
\bauthor{\bsnm{Clarkson}, \binits{C.}}:
\batitle{{Magnification and evolution biases in large-scale structure
  surveys}}.
\bjtitle{JCAP}
\bvolume{12}(\bissue{12}),
\bfpage{009}
(\byear{2021})
{\href{https://arxiv.org/abs/2107.13401}{{arXiv:2107.13401}}}
{[astro-ph.CO]}.
\doiurl{10.1088/1475-7516/2021/12/009}
\end{barticle}
\endbibitem

%%% 23
\bibitem{Fanizza:2022wob}
\begin{botherref}
\oauthor{\bsnm{Fanizza}, \binits{G.}},
\oauthor{\bsnm{Di~Dio}, \binits{E.}},
\oauthor{\bsnm{Durrer}, \binits{R.}},
\oauthor{\bsnm{Marozzi}, \binits{G.}}:
{The gauge invariant cosmological Jacobi map from weak lensing at leading
  order}
(2022)
{\href{https://arxiv.org/abs/2201.11552}{{arXiv:2201.11552}}}
{[astro-ph.CO]}
\end{botherref}
\endbibitem

%%% 24
\bibitem{Bartelmann:2010fz}
\begin{barticle}
\bauthor{\bsnm{Bartelmann}, \binits{M.}}:
\batitle{{Gravitational Lensing}}.
\bjtitle{Class. Quant. Grav.}
\bvolume{27},
\bfpage{233001}
(\byear{2010})
{\href{https://arxiv.org/abs/1010.3829}{{arXiv:1010.3829}}}
{[astro-ph.CO]}.
\doiurl{10.1088/0264-9381/27/23/233001}
\end{barticle}
\endbibitem

%%% 25
\bibitem{Montanari:2015rga}
\begin{barticle}
\bauthor{\bsnm{Montanari}, \binits{F.}},
\bauthor{\bsnm{Durrer}, \binits{R.}}:
\batitle{{Measuring the lensing potential with tomographic galaxy number
  counts}}.
\bjtitle{JCAP}
\bvolume{10},
\bfpage{070}
(\byear{2015})
{\href{https://arxiv.org/abs/1506.01369}{{arXiv:1506.01369}}}
{[astro-ph.CO]}.
\doiurl{10.1088/1475-7516/2015/10/070}
\end{barticle}
\endbibitem

%%% 26
\bibitem{Jelic-Cizmek:2020pkh}
\begin{barticle}
\bauthor{\bsnm{Jelic-Cizmek}, \binits{G.}},
\bauthor{\bsnm{Lepori}, \binits{F.}},
\bauthor{\bsnm{Bonvin}, \binits{C.}},
\bauthor{\bsnm{Durrer}, \binits{R.}}:
\batitle{{On the importance of lensing for galaxy clustering in photometric and
  spectroscopic surveys}}.
\bjtitle{JCAP}
\bvolume{04},
\bfpage{055}
(\byear{2021})
{\href{https://arxiv.org/abs/2004.12981}{{arXiv:2004.12981}}}
{[astro-ph.CO]}.
\doiurl{10.1088/1475-7516/2021/04/055}
\end{barticle}
\endbibitem

%%% 27
\bibitem{Euclid:2021rez}
\begin{botherref}
\oauthor{\bsnm{Lepori}, \binits{F.}}, et al.:
{Euclid preparation: XIX. Impact of magnification on photometric galaxy
  clustering}
(2021)
{\href{https://arxiv.org/abs/2110.05435}{{arXiv:2110.05435}}}
{[astro-ph.CO]}
\end{botherref}
\endbibitem

%%% 28
\bibitem{Menard:2002da}
\begin{barticle}
\bauthor{\bsnm{Menard}, \binits{B.}},
\bauthor{\bsnm{Bartelmann}, \binits{M.}},
\bauthor{\bsnm{Mellier}, \binits{Y.}}:
\batitle{{Measuring omega\_0 with higher-order quasar-galaxy correlations
  induced by weak lensing}}.
\bjtitle{Astron. Astrophys.}
\bvolume{409},
\bfpage{411}--\blpage{421}
(\byear{2003})
{\href{https://arxiv.org/abs/astro-ph/0208361}{{arXiv:astro-ph/0208361}}}.
\doiurl{10.1051/0004-6361:20031095}
\end{barticle}
\endbibitem

%%% 29
\bibitem{Kaiser1987}
\begin{barticle}
\bauthor{\bsnm{{Kaiser}}, \binits{N.}}:
\batitle{{Clustering in real space and in redshift space}}.
\bjtitle{M.N.R.A.S.}
\bvolume{227},
\bfpage{1}--\blpage{21}
(\byear{1987})
\end{barticle}
\endbibitem

%%% 30
\bibitem{BOSS:2016wmc}
\begin{barticle}
\bauthor{\bsnm{Alam}, \binits{S.}}, \betal:
\batitle{{The clustering of galaxies in the completed SDSS-III Baryon
  Oscillation Spectroscopic Survey: cosmological analysis of the DR12 galaxy
  sample}}.
\bjtitle{Mon. Not. Roy. Astron. Soc.}
\bvolume{470}(\bissue{3}),
\bfpage{2617}--\blpage{2652}
(\byear{2017})
{\href{https://arxiv.org/abs/1607.03155}{{arXiv:1607.03155}}}
{[astro-ph.CO]}.
\doiurl{10.1093/mnras/stx721}
\end{barticle}
\endbibitem

%%% 31
\bibitem{Beutler:2012px}
\begin{barticle}
\bauthor{\bsnm{Beutler}, \binits{F.}},
\bauthor{\bsnm{Blake}, \binits{C.}},
\bauthor{\bsnm{Colless}, \binits{M.}},
\bauthor{\bsnm{Jones}, \binits{D.H.}},
\bauthor{\bsnm{Staveley-Smith}, \binits{L.}},
\bauthor{\bsnm{Poole}, \binits{G.B.}},
\bauthor{\bsnm{Campbell}, \binits{L.}},
\bauthor{\bsnm{Parker}, \binits{Q.}},
\bauthor{\bsnm{Saunders}, \binits{W.}},
\bauthor{\bsnm{Watson}, \binits{F.}}:
\batitle{{The 6dF Galaxy Survey: $z \approx 0$ measurement of the growth rate
  and $\sigma_8$}}.
\bjtitle{Mon. Not. Roy. Astron. Soc.}
\bvolume{423},
\bfpage{3430}--\blpage{3444}
(\byear{2012})
{\href{https://arxiv.org/abs/1204.4725}{{arXiv:1204.4725}}}
{[astro-ph.CO]}.
\doiurl{10.1111/j.1365-2966.2012.21136.x}
\end{barticle}
\endbibitem

%%% 32
\bibitem{WiggleZ:2013akc}
\begin{barticle}
\bauthor{\bsnm{Contreras}, \binits{C.}}, \betal:
\batitle{{The WiggleZ Dark Energy Survey: measuring the cosmic growth rate with
  the two-point galaxy correlation function}}.
\bjtitle{Mon. Not. Roy. Astron. Soc.}
\bvolume{430},
\bfpage{924}
(\byear{2013})
{\href{https://arxiv.org/abs/1302.5178}{{arXiv:1302.5178}}}
{[astro-ph.CO]}.
\doiurl{10.1093/mnras/sts608}
\end{barticle}
\endbibitem

%%% 33
\bibitem{Samushia:2013yga}
\begin{barticle}
\bauthor{\bsnm{Samushia}, \binits{L.}}, \betal:
\batitle{{The clustering of galaxies in the SDSS-III Baryon Oscillation
  Spectroscopic Survey: measuring growth rate and geometry with anisotropic
  clustering}}.
\bjtitle{Mon. Not. Roy. Astron. Soc.}
\bvolume{439}(\bissue{4}),
\bfpage{3504}--\blpage{3519}
(\byear{2014})
{\href{https://arxiv.org/abs/1312.4899}{{arXiv:1312.4899}}}
{[astro-ph.CO]}.
\doiurl{10.1093/mnras/stu197}
\end{barticle}
\endbibitem

%%% 34
\bibitem{Reid:2014iaa}
\begin{barticle}
\bauthor{\bsnm{Reid}, \binits{B.A.}},
\bauthor{\bsnm{Seo}, \binits{H.-J.}},
\bauthor{\bsnm{Leauthaud}, \binits{A.}},
\bauthor{\bsnm{Tinker}, \binits{J.L.}},
\bauthor{\bsnm{White}, \binits{M.}}:
\batitle{{A 2.5 per cent measurement of the growth rate from small-scale
  redshift space clustering of SDSS-III CMASS galaxies}}.
\bjtitle{Mon. Not. Roy. Astron. Soc.}
\bvolume{444}(\bissue{1}),
\bfpage{476}--\blpage{502}
(\byear{2014})
{\href{https://arxiv.org/abs/1404.3742}{{arXiv:1404.3742}}}
{[astro-ph.CO]}.
\doiurl{10.1093/mnras/stu1391}
\end{barticle}
\endbibitem

%%% 35
\bibitem{BOSS:2016ntk}
\begin{barticle}
\bauthor{\bsnm{Satpathy}, \binits{S.}}, \betal:
\batitle{{The clustering of galaxies in the completed SDSS-III Baryon
  Oscillation Spectroscopic Survey: On the measurement of growth rate using
  galaxy correlation functions}}.
\bjtitle{Mon. Not. Roy. Astron. Soc.}
\bvolume{469}(\bissue{2}),
\bfpage{1369}--\blpage{1382}
(\byear{2017})
{\href{https://arxiv.org/abs/1607.03148}{{arXiv:1607.03148}}}
{[astro-ph.CO]}.
\doiurl{10.1093/mnras/stx883}
\end{barticle}
\endbibitem

%%% 36
\bibitem{Mohammad:2017lzz}
\begin{barticle}
\bauthor{\bsnm{Mohammad}, \binits{F.G.}}, \betal:
\batitle{{The VIMOS Public Extragalactic Redshift Survey (VIPERS). An unbiased
  estimate of the growth rate of structure at
  \ensuremath{\langle}z\ensuremath{\rangle} = 0.85 using the clustering of
  luminous blue galaxies}}.
\bjtitle{Astron. Astrophys.}
\bvolume{610},
\bfpage{59}
(\byear{2018})
{\href{https://arxiv.org/abs/1708.00026}{{arXiv:1708.00026}}}
{[astro-ph.CO]}.
\doiurl{10.1051/0004-6361/201731685}
\end{barticle}
\endbibitem

%%% 37
\bibitem{Shi:2017qpr}
\begin{barticle}
\bauthor{\bsnm{Shi}, \binits{F.}}, \betal:
\batitle{{Mapping the Real Space Distributions of Galaxies in SDSS DR7: II.
  Measuring the growth rate, clustering amplitude of matter and biases of
  galaxies at redshift $0.1$}}.
\bjtitle{Astrophys. J.}
\bvolume{861}(\bissue{2}),
\bfpage{137}
(\byear{2018})
{\href{https://arxiv.org/abs/1712.04163}{{arXiv:1712.04163}}}
{[astro-ph.CO]}.
\doiurl{10.3847/1538-4357/aacb20}
\end{barticle}
\endbibitem

%%% 38
\bibitem{Ruggeri:2018kdn}
\begin{barticle}
\bauthor{\bsnm{Ruggeri}, \binits{R.}}, \betal:
\batitle{{The clustering of the SDSS-IV extended Baryon Oscillation
  Spectroscopic Survey DR14 quasar sample: measuring the evolution of the
  growth rate using redshift space distortions between redshift 0.8 and 2.2}}.
\bjtitle{Mon. Not. Roy. Astron. Soc.}
\bvolume{483}(\bissue{3}),
\bfpage{3878}--\blpage{3887}
(\byear{2019})
{\href{https://arxiv.org/abs/1801.02891}{{arXiv:1801.02891}}}
{[astro-ph.CO]}.
\doiurl{10.1093/mnras/sty3395}
\end{barticle}
\endbibitem

%%% 39
\bibitem{Bautista:2020ahg}
\begin{barticle}
\bauthor{\bsnm{Bautista}, \binits{J.E.}}, \betal:
\batitle{{The Completed SDSS-IV extended Baryon Oscillation Spectroscopic
  Survey: measurement of the BAO and growth rate of structure of the luminous
  red galaxy sample from the anisotropic correlation function between redshifts
  0.6 and 1}}.
\bjtitle{Mon. Not. Roy. Astron. Soc.}
\bvolume{500}(\bissue{1}),
\bfpage{736}--\blpage{762}
(\byear{2020})
{\href{https://arxiv.org/abs/2007.08993}{{arXiv:2007.08993}}}
{[astro-ph.CO]}.
\doiurl{10.1093/mnras/staa2800}
\end{barticle}
\endbibitem

%%% 40
\bibitem{Lange:2021zre}
\begin{barticle}
\bauthor{\bsnm{Lange}, \binits{J.U.}},
\bauthor{\bsnm{Hearin}, \binits{A.P.}},
\bauthor{\bsnm{Leauthaud}, \binits{A.}},
\bauthor{\bparticle{van~den} \bsnm{Bosch}, \binits{F.C.}},
\bauthor{\bsnm{Guo}, \binits{H.}},
\bauthor{\bsnm{DeRose}, \binits{J.}}:
\batitle{{Five per\,cent measurements of the growth rate from simulation-based
  modelling of redshift-space clustering in BOSS LOWZ}}.
\bjtitle{Mon. Not. Roy. Astron. Soc.}
\bvolume{509}(\bissue{2}),
\bfpage{1779}--\blpage{1804}
(\byear{2021})
{\href{https://arxiv.org/abs/2101.12261}{{arXiv:2101.12261}}}
{[astro-ph.CO]}.
\doiurl{10.1093/mnras/stab3111}
\end{barticle}
\endbibitem

%%% 41
\bibitem{Bonvin:2013ogt}
\begin{barticle}
\bauthor{\bsnm{Bonvin}, \binits{C.}},
\bauthor{\bsnm{Hui}, \binits{L.}},
\bauthor{\bsnm{Gaztanaga}, \binits{E.}}:
\batitle{{Asymmetric galaxy correlation functions}}.
\bjtitle{Phys. Rev. D}
\bvolume{89}(\bissue{8}),
\bfpage{083535}
(\byear{2014})
{\href{https://arxiv.org/abs/1309.1321}{{arXiv:1309.1321}}}
{[astro-ph.CO]}.
\doiurl{10.1103/PhysRevD.89.083535}
\end{barticle}
\endbibitem

%%% 42
\bibitem{Gaztanaga:2015jrs}
\begin{barticle}
\bauthor{\bsnm{Gaztanaga}, \binits{E.}},
\bauthor{\bsnm{Bonvin}, \binits{C.}},
\bauthor{\bsnm{Hui}, \binits{L.}}:
\batitle{{Measurement of the dipole in the cross-correlation function of
  galaxies}}.
\bjtitle{JCAP}
\bvolume{01},
\bfpage{032}
(\byear{2017})
{\href{https://arxiv.org/abs/1512.03918}{{arXiv:1512.03918}}}
{[astro-ph.CO]}.
\doiurl{10.1088/1475-7516/2017/01/032}
\end{barticle}
\endbibitem

%%% 43
\bibitem{Bonvin:2018ckp}
\begin{barticle}
\bauthor{\bsnm{Bonvin}, \binits{C.}},
\bauthor{\bsnm{Fleury}, \binits{P.}}:
\batitle{{Testing the equivalence principle on cosmological scales}}.
\bjtitle{JCAP}
\bvolume{05},
\bfpage{061}
(\byear{2018})
{\href{https://arxiv.org/abs/1803.02771}{{arXiv:1803.02771}}}
{[astro-ph.CO]}.
\doiurl{10.1088/1475-7516/2018/05/061}
\end{barticle}
\endbibitem

%%% 44
\bibitem{Legrand:2020sek}
\begin{barticle}
\bauthor{\bsnm{Legrand}, \binits{L.}},
\bauthor{\bsnm{Hern\'andez-Monteagudo}, \binits{C.}},
\bauthor{\bsnm{Douspis}, \binits{M.}},
\bauthor{\bsnm{Aghanim}, \binits{N.}},
\bauthor{\bsnm{Angulo}, \binits{R.E.}}:
\batitle{{High resolution tomography for galaxy spectroscopic surveys with
  Angular Redshift Fluctuations}}.
\bjtitle{Astron. Astrophys.}
\bvolume{646},
\bfpage{109}
(\byear{2021})
{\href{https://arxiv.org/abs/2007.14412}{{arXiv:2007.14412}}}
{[astro-ph.CO]}.
\doiurl{10.1051/0004-6361/202039049}
\end{barticle}
\endbibitem

%%% 45
\bibitem{Matthewson:2022bpp}
\begin{botherref}
\oauthor{\bsnm{Matthewson}, \binits{W.}},
\oauthor{\bsnm{Stock}, \binits{D.}},
\oauthor{\bsnm{Durrer}, \binits{R.}}:
{Redshift weighted galaxy number counts}
(2022)
{\href{https://arxiv.org/abs/2203.07414}{{arXiv:2203.07414}}}
{[astro-ph.CO]}
\end{botherref}
\endbibitem

%%% 46
\bibitem{Chang:2007xk}
\begin{barticle}
\bauthor{\bsnm{Chang}, \binits{T.-C.}},
\bauthor{\bsnm{Pen}, \binits{U.-L.}},
\bauthor{\bsnm{Peterson}, \binits{J.B.}},
\bauthor{\bsnm{McDonald}, \binits{P.}}:
\batitle{{Baryon Acoustic Oscillation Intensity Mapping as a Test of Dark
  Energy}}.
\bjtitle{Phys. Rev. Lett.}
\bvolume{100},
\bfpage{091303}
(\byear{2008})
{\href{https://arxiv.org/abs/0709.3672}{{arXiv:0709.3672}}}
{[astro-ph]}.
\doiurl{10.1103/PhysRevLett.100.091303}
\end{barticle}
\endbibitem

%%% 47
\bibitem{Wyithe:2008mv}
\begin{barticle}
\bauthor{\bsnm{Wyithe}, \binits{S.}},
\bauthor{\bsnm{Loeb}, \binits{A.}}:
\batitle{{The 21cm Power Spectrum After Reionization}}.
\bjtitle{Mon. Not. Roy. Astron. Soc.}
\bvolume{397},
\bfpage{1926}
(\byear{2009})
{\href{https://arxiv.org/abs/0808.2323}{{arXiv:0808.2323}}}
{[astro-ph]}.
\doiurl{10.1111/j.1365-2966.2009.15019.x}
\end{barticle}
\endbibitem

%%% 48
\bibitem{Furlanetto:2006jb}
\begin{barticle}
\bauthor{\bsnm{Furlanetto}, \binits{S.}},
\bauthor{\bsnm{Oh}, \binits{S.P.}},
\bauthor{\bsnm{Briggs}, \binits{F.}}:
\batitle{{Cosmology at Low Frequencies: The 21 cm Transition and the
  High-Redshift Universe}}.
\bjtitle{Phys. Rept.}
\bvolume{433},
\bfpage{181}--\blpage{301}
(\byear{2006})
{\href{https://arxiv.org/abs/astro-ph/0608032}{{arXiv:astro-ph/0608032}}}.
\doiurl{10.1016/j.physrep.2006.08.002}
\end{barticle}
\endbibitem

%%% 49
\bibitem{10.1111/j.1745-3933.2008.00581.x}
\begin{barticle}
\bauthor{\bsnm{Pen}, \binits{U.-L.}},
\bauthor{\bsnm{Staveley-Smith}, \binits{L.}},
\bauthor{\bsnm{Peterson}, \binits{J.B.}},
\bauthor{\bsnm{Chang}, \binits{T.-C.}}:
\batitle{{First detection of cosmic structure in the 21-cm intensity field}}.
\bjtitle{Mon. Not. Roy. Astron. Soc. Lett.}
\bvolume{394}(\bissue{1}),
\bfpage{6}--\blpage{10}
(\byear{2009})
{\href{https://arxiv.org/abs/https://academic.oup.com/mnrasl/article-pdf/394/1/L6/3783109/394-1-L6.pdf}{{https://academic.oup.com/mnrasl/article-pdf/394/1/L6/3783109/394-1-L6.pdf}}}.
\doiurl{10.1111/j.1745-3933.2008.00581.x}
\end{barticle}
\endbibitem

%%% 50
\bibitem{Masui_2013}
\begin{barticle}
\bauthor{\bsnm{Masui}, \binits{K.W.}},
\bauthor{\bsnm{Switzer}, \binits{E.R.}},
\bauthor{\bsnm{Banavar}, \binits{N.}},
\bauthor{\bsnm{Bandura}, \binits{K.}},
\bauthor{\bsnm{Blake}, \binits{C.}},
\bauthor{\bsnm{Calin}, \binits{L.-M.}},
\bauthor{\bsnm{Chang}, \binits{T.-C.}},
\bauthor{\bsnm{Chen}, \binits{X.}},
\bauthor{\bsnm{Li}, \binits{Y.-C.}},
\bauthor{\bsnm{Liao}, \binits{Y.-W.}},
\bauthor{\bsnm{Natarajan}, \binits{A.}},
\bauthor{\bsnm{Pen}, \binits{U.-L.}},
\bauthor{\bsnm{Peterson}, \binits{J.B.}},
\bauthor{\bsnm{Shaw}, \binits{J.R.}},
\bauthor{\bsnm{Voytek}, \binits{T.C.}}:
\batitle{{Measurement} {of} 21 cm {Brightness} {Fluctuations} {at} z $\sim$ 0.8
  {in} {Cross}-{Correlation}}.
\bjtitle{Astrophys.~J.}
\bvolume{763}(\bissue{1}),
\bfpage{20}
(\byear{2013}).
\doiurl{10.1088/2041-8205/763/1/l20}
\end{barticle}
\endbibitem

%%% 51
\bibitem{Tramonte:2020csa}
\begin{barticle}
\bauthor{\bsnm{Tramonte}, \binits{D.}},
\bauthor{\bsnm{Ma}, \binits{Y.-Z.}}:
\batitle{{The neutral hydrogen distribution in large-scale haloes from 21-cm
  intensity maps}}.
\bjtitle{Mon. Not. Roy. Astron. Soc.}
\bvolume{498}(\bissue{4}),
\bfpage{5916}--\blpage{5935}
(\byear{2020})
{\href{https://arxiv.org/abs/2009.02387}{{arXiv:2009.02387}}}
{[astro-ph.CO]}.
\doiurl{10.1093/mnras/staa2727}
\end{barticle}
\endbibitem

%%% 52
\bibitem{Bowman:2018yin}
\begin{barticle}
\bauthor{\bsnm{Bowman}, \binits{J.D.}},
\bauthor{\bsnm{Rogers}, \binits{A.E.E.}},
\bauthor{\bsnm{Monsalve}, \binits{R.A.}},
\bauthor{\bsnm{Mozdzen}, \binits{T.J.}},
\bauthor{\bsnm{Mahesh}, \binits{N.}}:
\batitle{{An absorption profile centred at 78 megahertz in the sky-averaged
  spectrum}}.
\bjtitle{Nature}
\bvolume{555}(\bissue{7694}),
\bfpage{67}--\blpage{70}
(\byear{2018})
{\href{https://arxiv.org/abs/1810.05912}{{arXiv:1810.05912}}}
{[astro-ph.CO]}.
\doiurl{10.1038/nature25792}
\end{barticle}
\endbibitem

%%% 53
\bibitem{Kuhn:2021vgx}
\begin{botherref}
\oauthor{\bsnm{Kuhn}, \binits{E.R.}}, et al.:
{Design and implementation of a noise temperature measurement system for the
  Hydrogen Intensity and Real-time Analysis eXperiment (HIRAX)}
(2021)
{\href{https://arxiv.org/abs/2101.06337}{{arXiv:2101.06337}}}
{[astro-ph.IM]}.
\doiurl{10.1117/12.2560270}
\end{botherref}
\endbibitem

%%% 54
\bibitem{Yohana:2019ahg}
\begin{botherref}
\oauthor{\bsnm{Yohana}, \binits{E.}},
\oauthor{\bsnm{Li}, \binits{Y.-C.}},
\oauthor{\bsnm{Ma}, \binits{Y.-Z.}}:
{Forecasts of cosmological constraints from HI intensity mapping with FAST,
  BINGO \& SKA-I}
(2019)
{\href{https://arxiv.org/abs/1908.03024}{{arXiv:1908.03024}}}
{[astro-ph.CO]}.
\doiurl{10.1088/1674-4527/19/12/186}
\end{botherref}
\endbibitem

%%% 55
\bibitem{Barry:2021szi}
\begin{botherref}
\oauthor{\bsnm{Barry}, \binits{N.}},
\oauthor{\bsnm{Bernardi}, \binits{G.}},
\oauthor{\bsnm{Greig}, \binits{B.}},
\oauthor{\bsnm{Kern}, \binits{N.}},
\oauthor{\bsnm{Mertens}, \binits{F.}}:
{SKA-Low Intensity Mapping Pathfinder Updates: Deeper 21 cm Power Spectrum
  Limits from Improved Analysis Frameworks}
(2021)
{\href{https://arxiv.org/abs/2110.06173}{{arXiv:2110.06173}}}
{[astro-ph.CO]}
\end{botherref}
\endbibitem

%%% 56
\bibitem{CHIME:2022kvg}
\begin{botherref}
\oauthor{\bsnm{Amiri}, \binits{M.}}, et al.:
{Detection of Cosmological 21 cm Emission with the Canadian Hydrogen Intensity
  Mapping Experiment}
(2022)
{\href{https://arxiv.org/abs/2202.01242}{{arXiv:2202.01242}}}
{[astro-ph.CO]}
\end{botherref}
\endbibitem

%%% 57
\bibitem{Santos:2015gra}
\begin{barticle}
\bauthor{\bsnm{Santos}, \binits{M.G.}}, \betal:
\batitle{{Cosmology from a SKA HI intensity mapping survey}}.
\bjtitle{PoS}
\bvolume{AASKA14},
\bfpage{019}
(\byear{2015})
{\href{https://arxiv.org/abs/1501.03989}{{arXiv:1501.03989}}}
{[astro-ph.CO]}.
\doiurl{10.22323/1.215.0019}
\end{barticle}
\endbibitem

%%% 58
\bibitem{Abdalla:2015zra}
\begin{barticle}
\bauthor{\bsnm{Abdalla}, \binits{F.B.}}, \betal:
\batitle{{Cosmology from HI galaxy surveys with the SKA}}.
\bjtitle{PoS}
\bvolume{AASKA14},
\bfpage{017}
(\byear{2015}).
\doiurl{10.22323/1.215.0017}
\end{barticle}
\endbibitem

%%% 59
\bibitem{Fonseca:2018hsu}
\begin{barticle}
\bauthor{\bsnm{Fonseca}, \binits{J.}},
\bauthor{\bsnm{Maartens}, \binits{R.}},
\bauthor{\bsnm{Santos}, \binits{M.G.}}:
\batitle{{Synergies between intensity maps of hydrogen lines}}.
\bjtitle{Mon. Not. Roy. Astron. Soc.}
\bvolume{479}(\bissue{3}),
\bfpage{3490}--\blpage{3497}
(\byear{2018})
{\href{https://arxiv.org/abs/1803.07077}{{arXiv:1803.07077}}}
{[astro-ph.CO]}.
\doiurl{10.1093/mnras/sty1702}
\end{barticle}
\endbibitem

%%% 60
\bibitem{Hall:2012wd}
\begin{barticle}
\bauthor{\bsnm{Hall}, \binits{A.}},
\bauthor{\bsnm{Bonvin}, \binits{C.}},
\bauthor{\bsnm{Challinor}, \binits{A.}}:
\batitle{{Testing General Relativity with 21-cm intensity mapping}}.
\bjtitle{Phys. Rev. D}
\bvolume{87}(\bissue{6}),
\bfpage{064026}
(\byear{2013})
{\href{https://arxiv.org/abs/1212.0728}{{arXiv:1212.0728}}}
{[astro-ph.CO]}.
\doiurl{10.1103/PhysRevD.87.064026}
\end{barticle}
\endbibitem

%%% 61
\bibitem{Hall:2016bmm}
\begin{barticle}
\bauthor{\bsnm{Hall}, \binits{A.}},
\bauthor{\bsnm{Bonvin}, \binits{C.}}:
\batitle{{Measuring cosmic velocities with 21 cm intensity mapping and galaxy
  redshift survey cross-correlation dipoles}}.
\bjtitle{Phys. Rev. D}
\bvolume{95}(\bissue{4}),
\bfpage{043530}
(\byear{2017})
{\href{https://arxiv.org/abs/1609.09252}{{arXiv:1609.09252}}}
{[astro-ph.CO]}.
\doiurl{10.1103/PhysRevD.95.043530}
\end{barticle}
\endbibitem

%%% 62
\bibitem{Jalilvand:2019bhk}
\begin{barticle}
\bauthor{\bsnm{Jalilvand}, \binits{M.}},
\bauthor{\bsnm{Majerotto}, \binits{E.}},
\bauthor{\bsnm{Bonvin}, \binits{C.}},
\bauthor{\bsnm{Lacasa}, \binits{F.}},
\bauthor{\bsnm{Kunz}, \binits{M.}},
\bauthor{\bsnm{Naidoo}, \binits{W.}},
\bauthor{\bsnm{Moodley}, \binits{K.}}:
\batitle{{New Estimator for Gravitational Lensing Using Galaxy and Intensity
  Mapping Surveys}}.
\bjtitle{Phys. Rev. Lett.}
\bvolume{124}(\bissue{3}),
\bfpage{031101}
(\byear{2020})
{\href{https://arxiv.org/abs/1907.00071}{{arXiv:1907.00071}}}
{[astro-ph.CO]}.
\doiurl{10.1103/PhysRevLett.124.031101}
\end{barticle}
\endbibitem

%%% 63
\bibitem{Kovetz:2019uss}
\begin{barticle}
\bauthor{\bsnm{Kovetz}, \binits{E.D.}}, \betal:
\batitle{{Astrophysics and Cosmology with Line-Intensity Mapping}}.
\bjtitle{Bull. Am. Astron. Soc.}
\bvolume{51}(\bissue{3}),
\bfpage{101}
(\byear{2020})
{\href{https://arxiv.org/abs/1903.04496}{{arXiv:1903.04496}}}
{[astro-ph.CO]}
\end{barticle}
\endbibitem

%%% 64
\bibitem{Perlick:2004tq}
\begin{barticle}
\bauthor{\bsnm{Perlick}, \binits{V.}}:
\batitle{{Gravitational lensing from a spacetime perspective}}.
\bjtitle{Living Rev. Rel.}
\bvolume{7},
\bfpage{9}
(\byear{2004})
\end{barticle}
\endbibitem

%%% 65
\bibitem{Bartelmann:1999yn}
\begin{barticle}
\bauthor{\bsnm{Bartelmann}, \binits{M.}},
\bauthor{\bsnm{Schneider}, \binits{P.}}:
\batitle{{Weak gravitational lensing}}.
\bjtitle{Phys. Rept.}
\bvolume{340},
\bfpage{291}--\blpage{472}
(\byear{2001})
{\href{https://arxiv.org/abs/astro-ph/9912508}{{arXiv:astro-ph/9912508}}}.
\doiurl{10.1016/S0370-1573(00)00082-X}
\end{barticle}
\endbibitem

%%% 66
\bibitem{Kilbinger:2014cea}
\begin{barticle}
\bauthor{\bsnm{Kilbinger}, \binits{M.}}:
\batitle{{Cosmology with cosmic shear observations: a review}}.
\bjtitle{Rept. Prog. Phys.}
\bvolume{78},
\bfpage{086901}
(\byear{2015})
{\href{https://arxiv.org/abs/1411.0115}{{arXiv:1411.0115}}}
{[astro-ph.CO]}.
\doiurl{10.1088/0034-4885/78/8/086901}
\end{barticle}
\endbibitem

%%% 67
\bibitem{Hildebrandt:2018yau}
\begin{barticle}
\bauthor{\bsnm{Hildebrandt}, \binits{H.}}, \betal:
\batitle{{KiDS+VIKING-450: Cosmic shear tomography with optical and infrared
  data}}.
\bjtitle{Astron. Astrophys.}
\bvolume{633},
\bfpage{69}
(\byear{2020})
{\href{https://arxiv.org/abs/1812.06076}{{arXiv:1812.06076}}}
{[astro-ph.CO]}.
\doiurl{10.1051/0004-6361/201834878}
\end{barticle}
\endbibitem

%%% 68
\bibitem{KiDS:2020suj}
\begin{barticle}
\bauthor{\bsnm{Asgari}, \binits{M.}}, \betal:
\batitle{{KiDS-1000 Cosmology: Cosmic shear constraints and comparison between
  two point statistics}}.
\bjtitle{Astron. Astrophys.}
\bvolume{645},
\bfpage{104}
(\byear{2021})
{\href{https://arxiv.org/abs/2007.15633}{{arXiv:2007.15633}}}
{[astro-ph.CO]}.
\doiurl{10.1051/0004-6361/202039070}
\end{barticle}
\endbibitem

%%% 69
\bibitem{DES:2020daw}
\begin{barticle}
\bauthor{\bsnm{Doux}, \binits{C.}}, \betal:
\batitle{{Consistency of cosmic shear analyses in harmonic and real space}}.
\bjtitle{Mon. Not. Roy. Astron. Soc.}
\bvolume{503}(\bissue{3}),
\bfpage{3796}--\blpage{3817}
(\byear{2021})
{\href{https://arxiv.org/abs/2011.06469}{{arXiv:2011.06469}}}
{[astro-ph.CO]}.
\doiurl{10.1093/mnras/stab661}
\end{barticle}
\endbibitem

%%% 70
\bibitem{DES:2021wwk}
\begin{botherref}
\oauthor{\bsnm{Abbott}, \binits{T.M.C.}}, et al.:
{Dark Energy Survey Year 3 Results: Cosmological Constraints from Galaxy
  Clustering and Weak Lensing}
(2021)
{\href{https://arxiv.org/abs/2105.13549}{{arXiv:2105.13549}}}
{[astro-ph.CO]}
\end{botherref}
\endbibitem

%%% 71
\bibitem{Amon:2022ycy}
\begin{botherref}
\oauthor{\bsnm{Amon}, \binits{A.}}, et al.:
{Consistent lensing and clustering in a low-$S_8$ Universe with BOSS, DES Year
  3, HSC Year 1 and KiDS-1000}
(2022)
{\href{https://arxiv.org/abs/2202.07440}{{arXiv:2202.07440}}}
{[astro-ph.CO]}
\end{botherref}
\endbibitem

%%% 72
\bibitem{DES:2022qpf}
\begin{botherref}
\oauthor{\bsnm{Doux}, \binits{C.}}, et al.:
{Dark Energy Survey Year 3 results: cosmological constraints from the analysis
  of cosmic shear in harmonic space}
(2022)
{\href{https://arxiv.org/abs/2203.07128}{{arXiv:2203.07128}}}
{[astro-ph.CO]}
\end{botherref}
\endbibitem

%%% 73
\bibitem{Kristian:1965sz}
\begin{barticle}
\bauthor{\bsnm{Kristian}, \binits{J.}},
\bauthor{\bsnm{Sachs}, \binits{R.K.}}:
\batitle{{Observations in cosmology}}.
\bjtitle{Astrophys. J.}
\bvolume{143},
\bfpage{379}--\blpage{399}
(\byear{1966}).
\doiurl{10.1086/148522}
\end{barticle}
\endbibitem

%%% 74
\bibitem{1983ApJ...271..431V}
\begin{barticle}
\bauthor{\bsnm{{Valdes}}, \binits{F.}},
\bauthor{\bsnm{{Tyson}}, \binits{J.A.}},
\bauthor{\bsnm{{Jarvis}}, \binits{J.F.}}:
\batitle{{Alignment of faint galaxy images : cosmological distortion and
  rotation.}}
\bjtitle{Astrophys. J.}
\bvolume{271},
\bfpage{431}--\blpage{441}
(\byear{1983}).
\doiurl{10.1086/161210}
\end{barticle}
\endbibitem

%%% 75
\bibitem{Francfort:2022laa}
\begin{botherref}
\oauthor{\bsnm{Francfort}, \binits{J.}},
\oauthor{\bsnm{Cusin}, \binits{G.}},
\oauthor{\bsnm{Durrer}, \binits{R.}}:
{A new observable for cosmic shear}
(2022)
{\href{https://arxiv.org/abs/2203.13634}{{arXiv:2203.13634}}}
{[astro-ph.CO]}
\end{botherref}
\endbibitem

%%% 76
\bibitem{Stebbins:1996wx}
\begin{botherref}
\oauthor{\bsnm{Stebbins}, \binits{A.}}:
{Weak lensing on the celestial sphere}
(1996)
{\href{https://arxiv.org/abs/astro-ph/9609149}{{arXiv:astro-ph/9609149}}}
\end{botherref}
\endbibitem

%%% 77
\bibitem{Schmidt:2012ne}
\begin{barticle}
\bauthor{\bsnm{Schmidt}, \binits{F.}},
\bauthor{\bsnm{Jeong}, \binits{D.}}:
\batitle{{Cosmic Rulers}}.
\bjtitle{Phys. Rev. D}
\bvolume{86},
\bfpage{083527}
(\byear{2012})
{\href{https://arxiv.org/abs/1204.3625}{{arXiv:1204.3625}}}
{[astro-ph.CO]}.
\doiurl{10.1103/PhysRevD.86.083527}
\end{barticle}
\endbibitem

%%% 78
\bibitem{Pranav:2018pnu}
\begin{barticle}
\bauthor{\bsnm{Pranav}, \binits{P.}},
\bauthor{\bparticle{van~de} \bsnm{Weygaert}, \binits{R.}},
\bauthor{\bsnm{Vegter}, \binits{G.}},
\bauthor{\bsnm{Jones}, \binits{B.J.T.}},
\bauthor{\bsnm{Adler}, \binits{R.J.}},
\bauthor{\bsnm{Feldbrugge}, \binits{J.}},
\bauthor{\bsnm{Park}, \binits{C.}},
\bauthor{\bsnm{Buchert}, \binits{T.}},
\bauthor{\bsnm{Kerber}, \binits{M.}}:
\batitle{{Topology and Geometry of Gaussian random fields I: on Betti Numbers,
  Euler characteristic and Minkowski functionals}}.
\bjtitle{Mon. Not. Roy. Astron. Soc.}
\bvolume{485}(\bissue{3}),
\bfpage{4167}--\blpage{4208}
(\byear{2019})
{\href{https://arxiv.org/abs/1812.07310}{{arXiv:1812.07310}}}
{[astro-ph.CO]}.
\doiurl{10.1093/mnras/stz541}
\end{barticle}
\endbibitem

%%% 79
\bibitem{Liu:2022vtr}
\begin{botherref}
\oauthor{\bsnm{Liu}, \binits{W.}},
\oauthor{\bsnm{Jiang}, \binits{A.}},
\oauthor{\bsnm{Fang}, \binits{W.}}:
{Probing massive neutrinos with the Minkowski functionals of large-scale
  structure}
(2022)
{\href{https://arxiv.org/abs/2204.02945}{{arXiv:2204.02945}}}
{[astro-ph.CO]}
\end{botherref}
\endbibitem

%%% 80
\bibitem{DiDio:2018unb}
\begin{barticle}
\bauthor{\bsnm{Di~Dio}, \binits{E.}},
\bauthor{\bsnm{Durrer}, \binits{R.}},
\bauthor{\bsnm{Maartens}, \binits{R.}},
\bauthor{\bsnm{Montanari}, \binits{F.}},
\bauthor{\bsnm{Umeh}, \binits{O.}}:
\batitle{{The Full-Sky Angular Bispectrum in Redshift Space}}.
\bjtitle{JCAP}
\bvolume{04},
\bfpage{053}
(\byear{2019})
{\href{https://arxiv.org/abs/1812.09297}{{arXiv:1812.09297}}}
{[astro-ph.CO]}.
\doiurl{10.1088/1475-7516/2019/04/053}
\end{barticle}
\endbibitem

%%% 81
\bibitem{deWeerd:2019cae}
\begin{barticle}
\bauthor{\bparticle{de} \bsnm{Weerd}, \binits{E.M.}},
\bauthor{\bsnm{Clarkson}, \binits{C.}},
\bauthor{\bsnm{Jolicoeur}, \binits{S.}},
\bauthor{\bsnm{Maartens}, \binits{R.}},
\bauthor{\bsnm{Umeh}, \binits{O.}}:
\batitle{{Multipoles of the relativistic galaxy bispectrum}}.
\bjtitle{JCAP}
\bvolume{05},
\bfpage{018}
(\byear{2020})
{\href{https://arxiv.org/abs/1912.11016}{{arXiv:1912.11016}}}
{[astro-ph.CO]}.
\doiurl{10.1088/1475-7516/2020/05/018}
\end{barticle}
\endbibitem

%%% 82
\bibitem{Durrer:2020orn}
\begin{barticle}
\bauthor{\bsnm{Durrer}, \binits{R.}},
\bauthor{\bsnm{Jalilvand}, \binits{M.}},
\bauthor{\bsnm{Kothari}, \binits{R.}},
\bauthor{\bsnm{Maartens}, \binits{R.}},
\bauthor{\bsnm{Montanari}, \binits{F.}}:
\batitle{{Full-sky bispectrum in redshift space for 21cm intensity maps}}.
\bjtitle{JCAP}
\bvolume{12},
\bfpage{003}
(\byear{2020})
{\href{https://arxiv.org/abs/2008.02266}{{arXiv:2008.02266}}}
{[astro-ph.CO]}.
\doiurl{10.1088/1475-7516/2020/12/003}
\end{barticle}
\endbibitem

%%% 83
\bibitem{Bermejo-Climent:2021jxf}
\begin{barticle}
\bauthor{\bsnm{Bermejo-Climent}, \binits{J.R.}},
\bauthor{\bsnm{Ballardini}, \binits{M.}},
\bauthor{\bsnm{Finelli}, \binits{F.}},
\bauthor{\bsnm{Paoletti}, \binits{D.}},
\bauthor{\bsnm{Maartens}, \binits{R.}},
\bauthor{\bsnm{Rubi\~no-Mart\'\i{}n}, \binits{J.A.}},
\bauthor{\bsnm{Valenziano}, \binits{L.}}:
\batitle{{Cosmological parameter forecasts by a joint 2D tomographic approach
  to CMB and galaxy clustering}}.
\bjtitle{Phys. Rev. D}
\bvolume{103}(\bissue{10}),
\bfpage{103502}
(\byear{2021})
{\href{https://arxiv.org/abs/2106.05267}{{arXiv:2106.05267}}}
{[astro-ph.CO]}.
\doiurl{10.1103/PhysRevD.103.103502}
\end{barticle}
\endbibitem

%%% 84
\bibitem{Riess:2021jrx}
\begin{botherref}
\oauthor{\bsnm{Riess}, \binits{A.G.}}, et al.:
{A Comprehensive Measurement of the Local Value of the Hubble Constant with 1
  km/s/Mpc Uncertainty from the Hubble Space Telescope and the SH0ES Team}
(2021)
{\href{https://arxiv.org/abs/2112.04510}{{arXiv:2112.04510}}}
{[astro-ph.CO]}
\end{botherref}
\endbibitem

\end{thebibliography}


\end{document}